\documentclass[runningheads]{llncs}
\usepackage{graphicx}
\usepackage{amsmath}
\usepackage{amssymb}
\usepackage{mathtools}
\usepackage{hyperref}
\usepackage{subcaption}
\usepackage{xcolor}
\makeatletter
\newcounter{tmp@cnt}
\newcommand*\@labelpunc{.}
\newcommand*\combine[1][2]{%
    \refstepcounter{enumi}
    \setcounter{tmp@cnt}{\value{enumi}}
    \addtocounter{enumi}{#1-1}
    \item[\thetmp@cnt--\theenumi\@labelpunc]}
\newcommand*\labeltype[2][]{\gdef\@labelpunc{#1}\renewcommand\thetmp@cnt{#2{tmp@cnt}}}
\makeatother

\allowdisplaybreaks

\DeclareMathOperator{\Dom}{Dom}
\DeclareMathOperator{\Rng}{Rng}
\DeclareMathOperator{\Proj}{Proj}
\DeclareMathOperator{\defined}{!}

\DeclareMathOperator{\Sig}{Sig}
\DeclareMathOperator{\level}{level}

\begin{document}
\title{Composition and Weight Pushing of Monotonic Subsequential Failure Transducers Representing Probabilistic Models}
\titlerunning{Constructions for Monotonic Subsequential Failure Transducers}
%
\author{Diana Geneva \and
Georgi Shopov \and
Stoyan Mihov}
\authorrunning{D. Geneva et al.}
%
\institute{Institute of Information and Communication Technologies\\
Bulgarian Academy of Sciences\\
2, Acad. G. Bonchev Str., 1113 Sofia, Bulgaria\\
\email{\{dageneva,gshopov,stoyan\}@lml.bas.bg}}
\maketitle              
\begin{abstract}
	We present a construction for the composition of subsequential transducers (representing conditional probabilistic models) with subsequential failure transducers (representing probabilistic models). Under certain conditions, satisfied by the corresponding transduction devices, a more efficient construction is applicable that avoids the creation of unnecessary states. Furthermore, the weights of the resulting failure transducers can be efficiently redistributed via weight pushing in the \(\langle \mathbb{R}_+, +, \times, 0, 1 \rangle\) and \(\langle \mathbb{R}_+, \max, \times, 0, 1 \rangle\) semirings.
\end{abstract}

\section{Introduction}
Failure transducers are widely used for representing n-gram back-off language models \cite{Allauzen2003GeneralizedAF}. Often the language model is composed with a given conditional probability distribution represented as a weighted finite-state transducer. However, usually, the transduction device used in the composition to represent the language model is not a failure transducer but rather a weighted transducer in which the back-off transitions are represented by epsilon transitions \cite{Mohri2008}.

In this paper formal constructions for performing composition and weight pushing of failure finite-state transducers are presented together with their corresponding correctness proofs. The obtained results enable the development of efficient implementations of the considered operations, which facilitate the practical application of failure weighted finite-state transducers for language modelling.

\section{Formal Preliminaries}\label{sec-prelim}

We begin by listing some standard notions that we use. An \emph{alphabet} is a finite set of symbols. Words of length \(n \ge 0\) over an alphabet \(\Sigma\) are introduced as usual and written \(a_1 a_2 \ldots a_n\), where \((\forall 1 \le i \le n)(a_i \in \Sigma)\). The unique word of length \(0\) is written \(\varepsilon\). The \emph{concatenation} of two words \(\alpha \coloneqq a_1 a_2 \ldots a_n\) and \(\beta \coloneqq b_1 b_2 \ldots b_m\) is \(\alpha\beta \coloneqq a_1 a_2 \ldots a_n b_1 b_2 \ldots b_m\). \(\Sigma^*\) denotes the set of all words over the alphabet \(\Sigma\). A \emph{language} over \(\Sigma\) is any subset of \(\Sigma^*\). A language \(L\) over \(\Sigma\) is \emph{prefix-free} if \((\forall \alpha, \beta \in L)((\exists \gamma \in \Sigma^*)(\alpha\gamma = \beta) \implies \alpha = \beta)\).

\begin{definition}\label{monoid-definition}
	A \emph{monoid} is a tuple \(\langle M, \otimes, \bar{1} \rangle\), where \(M\) is a set, \(\otimes \colon M \to M\) is a total associative function, i.e. \((\forall a, b, c \in M)((a \otimes b) \otimes c = a \otimes (b \otimes c))\), and \(\bar{1} \in M\) is the unit element, i.e. \((\forall a \in M)(a \otimes \bar{1} = \bar{1} \otimes a = a)\). If \(\otimes\) is commutative, i.e. \((\forall a, b \in M)(a \otimes b = b \otimes a)\), then the monoid is \emph{commutative}.
\end{definition}

The set \(\Sigma^*\) with concatenation as monoid operation and \(\varepsilon\) as unit element is a monoid (also denoted \(\Sigma^*\)) called the \emph{free monoid} over \(\Sigma\).

\begin{definition}
	The \emph{cartesian product} of the monoids \(\mathcal{M} \coloneqq \langle M, \otimes_M, \bar{1}_M \rangle\) and \(\mathcal{N} \coloneqq \langle N, \otimes_N, \bar{1}_N \rangle\) is \(\mathcal{M} \times \mathcal{N} \coloneqq \langle M \times N, \otimes_{M \times N}, \langle \bar{1}_M, \bar{1}_N \rangle \rangle\), where \(\otimes_{M \times N} \colon (M \times N)^2 \to M \times N\) is defined as \(\langle a_1, a_2 \rangle \otimes_{M \times N} \langle b_1, b_2 \rangle \coloneqq \langle a_1 \otimes_M b_1, a_2 \otimes_N b_2 \rangle\).
\end{definition}

\begin{remark}
	It can be easily verified that the cartesian product of the monoids \(\mathcal{M}\) and \(\mathcal{N}\) satisfies the conditions of \autoref{monoid-definition} and thus \(\mathcal{M} \times \mathcal{N}\) is a monoid. When no ambiguity occurs, we will use \(\odot\) to denote the monoid operation of \(\mathcal{M} \times \mathcal{N}\).
\end{remark}

In this paper we consider only subsequential transduction devices introduced by Sch\"utzenberger \cite{SCHUTZENBERGER197747}, which is why, for reasons of brevity, we will deliberately omit the word ``subsequential''.

\begin{definition}
	A \emph{transducer} is a tuple \(\langle \Sigma, \langle M, \otimes, \bar{1} \rangle, Q, s, F, \delta, \lambda, \iota, \rho \rangle\), where \(\Sigma\) is an alphabet, \(\langle M, \otimes, \bar{1} \rangle\) is a monoid, \(Q\) is a finite set of states, \(s \in Q\) is an initial state, \(F \subseteq Q\) is a set of final states, \(\delta \colon Q \times \Sigma \to Q\) is a partial transition function, \(\lambda \colon Q \times \Sigma \to M\) is a partial output function, \(\iota \in M\) is an initial output, \(\rho \colon F \to M\) is a total final output function, and \(\Dom(\delta) = \Dom(\lambda)\).
\end{definition}

\begin{definition}
	Let \(\langle \Sigma, \langle M, \otimes, \bar{1} \rangle, Q, s, F, \delta, \lambda, \iota, \rho \rangle\) be a transducer. The \emph{generalized transition function} \(\delta^* \colon Q \times \Sigma^* \to Q\) and the \emph{generalized output function} \(\lambda^* \colon Q \times \Sigma^* \to M\) are defined inductively as
	\begin{itemize}
		\item \(\delta^*(q, \varepsilon) \coloneqq q\) and \(\lambda^*(q, \varepsilon) \coloneqq \bar{1}\) for every \(q \in Q\);
		\item \(\delta^*(q, \alpha a) \coloneqq \delta(\delta^*(q, \alpha), a)\) and \(\lambda^*(q, \alpha a) \coloneqq \lambda^*(q, \alpha) \otimes \lambda(\delta^*(q, \alpha), a)\) for every \(q \in Q\), \(\alpha \in \Sigma^*\) and \(a \in \Sigma\).
	\end{itemize}
\end{definition}

\begin{definition}
	Let \(\mathcal{T} \coloneqq \langle \Sigma, \langle M, \otimes, \bar{1} \rangle, Q, s, F, \delta, \lambda, \iota, \rho \rangle\) be a transducer. For each \(q \in Q\) the function \(\mathcal{O}_\mathcal{T}^q \colon \Sigma^* \to M\) is defined for \(\alpha \in \Sigma^*\) as
	\[\mathcal{O}_\mathcal{T}^q(\alpha) \coloneqq
	  \begin{cases}
	  	\lambda^*(q, \alpha) \otimes \rho(\delta^*(q, \alpha)) & \text{if } \delta^*(q, \alpha) \in F \\
		\neg \defined & \text{otherwise}
	  \end{cases}\]
	The function \(\mathcal{O}_\mathcal{T} \colon \Sigma^* \to M\), defined for \(\alpha \in \Sigma^*\) as \(\mathcal{O}_\mathcal{T}(\alpha) \coloneqq \iota \otimes \mathcal{O}_\mathcal{T}^s(\alpha),\) is \emph{the function represented by the transducer \(\mathcal{T}\)}. A state \(q \in Q\) is \emph{co-accessible} in \(\mathcal{T}\) if \(\Dom(\mathcal{O}_\mathcal{T}^q) \neq \varnothing\).
\end{definition}

With \(\mathcal{R}\) we will denote the monoid \(\langle \mathbb{R}_+, \times, 1 \rangle\) of non-negative real numbers with multiplication as the monoid operation.

\begin{definition}
	Let \(\mathcal{T} \coloneqq \langle \Sigma, \mathcal{R}, Q, s, F, \delta, \lambda, \iota, \rho \rangle\) be a transducer. \(\mathcal{T}\) is \emph{probabilistic} if \(\mathcal{O}_\mathcal{T}\) is a probability distribution over \(\Sigma^*\), i.e. \((\forall \alpha \in \Dom(\mathcal{O}_\mathcal{T}))(\mathcal{O}_\mathcal{T}(\alpha) \in [0, 1])\) and \(\sum_{\alpha \in \Dom(\mathcal{O}_\mathcal{T})} \mathcal{O}_\mathcal{T}(\alpha) = 1\).
\end{definition}

Given a transducer \(\langle \Sigma, \mathcal{R}, Q, s, F, \delta, \lambda, \iota, \rho \rangle\), we will use the expression \(e(q)\) to mean \(\rho(q)\) if \(q \in F\) and \(0\) otherwise.

\begin{definition}
	Let \(\mathcal{T} \coloneqq \langle \Sigma, \mathcal{R}, Q, s, F, \delta, \lambda, \iota, \rho \rangle\) be a probabilistic transducer. \(\mathcal{T}\) is \emph{stochastic} if
		\[(\forall q \in Q)\left(e(q) + \sum_{a \in \Sigma \colon \defined\delta(q, a)} \lambda(q, a) = 1\right).\]
\end{definition}

\begin{definition}
	Let \(\mathcal{T} \coloneqq \langle \Sigma, \Omega^* \times \langle M, \otimes, \bar{1} \rangle, Q, s, F, \delta, \lambda, \iota, \rho \rangle\) be a transducer. For every \(q \in Q\) and \(\beta \in \Omega^*\) we define the function \(\mathcal{O}_\mathcal{T}^p(\bullet \mid \beta) \colon \Sigma^* \to M\) for \(\alpha \in \Sigma^*\) as
	\[\mathcal{O}_\mathcal{T}^p(\alpha \mid \beta) \coloneqq
	  \begin{cases}
	  	m & \text{if } \mathcal{O}_\mathcal{T}^p(\alpha) = \langle \beta, m \rangle \\
		\neg\defined & \text{otherwise}
	  \end{cases}\]
	  Moreover, for every \(\beta \in \Omega^*\) the function \(\mathcal{O}_\mathcal{T}(\bullet \mid \beta) \colon \Sigma^* \to M\) is defined for \(\alpha \in \Sigma^*\) as \(\mathcal{O}_\mathcal{T}(\alpha \mid \beta) \coloneqq \iota \otimes \mathcal{O}_\mathcal{T}^s(\alpha \mid \beta)\).
\end{definition}

\begin{definition}
	A \emph{conditional probabilistic transducer} is a transducer \(\mathcal{T} \coloneqq \langle \Sigma,\allowbreak \Omega^* \times \mathcal{R}, Q, s, F, \delta, \lambda, \iota, \rho \rangle\) such that
	\[(\forall \beta \in \Proj_1(\Rng(\mathcal{O}_\mathcal{T})))\left(\sum_{\alpha \in \Dom(\mathcal{O}_\mathcal{T}(\bullet \mid \beta))} \mathcal{O}_\mathcal{T}(\alpha \mid \beta) = 1\right).\]
\end{definition}

\begin{definition}
	A \emph{semiring} is a tuple \(\langle S, \oplus, \otimes, \bar{0}, \bar{1} \rangle\), where \(\langle S, \oplus, \bar{0} \rangle\) is a commutative monoid, \(\langle S, \otimes, \bar{1} \rangle\) is a monoid, \(\otimes\) distributes over \(\oplus\), i.e. \((\forall a, b, c \in S)(a \otimes (b \oplus c) = (a \otimes b) \oplus (a \otimes c) \land (a \oplus b) \otimes c = (a \otimes c) \oplus (b \otimes  c))\), and \(\bar{0}\) acts as an annihilator, i.e. \((\forall a \in S)(a \otimes \bar{0} = \bar{0} \otimes a = \bar{0})\).
\end{definition}

In the context of a semiring \(\langle S, \otimes, \oplus, \bar{0}, \bar{1} \rangle\) and an infinite set \(X \subseteq S\) indexed by \(I\), when we use the notation \(\bigoplus_{i \in I} x_i\) we will assume that the infinite sum is defined in this semiring and that it has the following properties:
\begin{itemize}
	\item \(\bigoplus_{i_1 \in I_1} x_{i_1} \oplus \bigoplus_{i_2 \in I_2} x_{i_2} = \bigoplus_{i \in I} x_i\) for every partition \(I_1, I_2\) of \(I\),
	\item \(s \otimes \bigoplus_{i \in I} x_i = \bigoplus_{i \in I} s \otimes x_i\) for every \(s \in S\).
\end{itemize}

\(\mathcal{R}^{\max} \coloneqq \langle \mathbb{R}_+, \max, \times, 0, 1 \rangle\) and \(\mathcal{R}^+ \coloneqq \langle \mathbb{R}_+, +, \times, 0, 1 \rangle\) are semirings. Infinite sums in \(\mathcal{R}^{\max}\) and \(\mathcal{R}^+\) are defined respectively as the supremum and the sum of the series.

\begin{definition}
	Let \(\mathcal{T} \coloneqq \langle \Sigma, \langle S, \otimes, \bar{1} \rangle, Q, s, F, \delta, \lambda, \iota, \rho \rangle\) be a transducer and \(\mathcal{S} \coloneqq \langle S, \oplus, \otimes, \bar{0}, \bar{1} \rangle\) be a semiring. \(\mathcal{T}\) is \emph{canonical} with respect to \(\mathcal{S}\) if
		\[(\forall q \in Q)\left(\bigoplus_{\alpha \in \Dom(\mathcal{O}_\mathcal{T}^q)} \mathcal{O}_\mathcal{T}^q(\alpha) = \bar{1}\right).\]
\end{definition}

We extend the expression \(e(q)\) for \(q \in Q\) in the context of a transducer \(\langle \Sigma, \langle S, \otimes, \bar{1} \rangle, Q,\allowbreak s, F, \delta, \lambda, \iota, \rho \rangle\) and a semiring \(\langle S, \oplus, \otimes, \bar{0}, \bar{1} \rangle\) to mean \(\rho(q)\) if \(q \in F\) and \(\bar{0}\) otherwise.

\begin{proposition}\label{canonical-property}
	Let \(\mathcal{T} \coloneqq \langle \Sigma, \langle S, \otimes, \bar{1} \rangle, Q, s, F, \delta, \lambda, \iota, \rho \rangle\) be a canonical transducer with respect to \(\langle S, \oplus, \otimes, \bar{0}, \bar{1} \rangle\). Then
	\[(\forall q \in Q)\left(e(q) \oplus \bigoplus_{a \in \Sigma \colon \defined \delta(q, a)} \lambda(q, a) = \bar{1}\right).\]
	\begin{proof}
		Follows directly from the fact that for every \(q \in Q\)
		\begin{align*}
			\bar{1} &= \bigoplus_{\alpha \in \Dom(\mathcal{O}_\mathcal{T}^q)} \mathcal{O}_\mathcal{T}^q(\alpha) \\
			&= e(q) \oplus \bigoplus_{\alpha \in \Dom(\mathcal{O}_\mathcal{T}^q) \setminus \{\varepsilon\}} \mathcal{O}_\mathcal{T}^q(\alpha) \\
			&= e(q) \oplus \bigoplus_{a \in \Sigma \colon \defined \delta(q, a)} \left(\lambda(q, a) \otimes \bigoplus_{\alpha \in \Dom(\mathcal{O}_\mathcal{T}^{\delta(q, a)})} \mathcal{O}_\mathcal{T}^{\delta(q, a)}(\alpha)\right) \\
			&= e(q) \oplus \bigoplus_{a \in \Sigma \colon \defined \delta(q, a)} \lambda(q, a).
		\end{align*}
	\end{proof}
\end{proposition}

It follows that every probabilistic transducer, which is canonical with respect to \(\mathcal{R}^+\), is stochastic. For a stochastic transducer the opposite is also true.

\begin{proposition}\label{stochastic-property}
	Let \(\mathcal{T} \coloneqq \langle \Sigma, \mathcal{R}, Q, s, F, \delta, \lambda, \iota, \rho \rangle\) be a stochastic transducer. Then \(\mathcal{T}\) is canonical with respect to \(\mathcal{R}^+\).
\end{proposition}

\begin{definition}
	A \emph{failure transducer} is a tuple \(\langle \Sigma, \langle M, \otimes, \bar{1} \rangle, Q, s, F, \delta, \lambda, \iota, \rho, f,\allowbreak \varphi \rangle\) where \(\langle \Sigma, \langle M, \otimes, \bar{1} \rangle, Q, s, F, \delta, \lambda, \iota, \rho \rangle\) is a transducer, \(f \colon Q \to Q\) is a partial failure transition function, \(\varphi \colon Q \to M\) is a partial failure output function, and \(\Dom(f) = \Dom(\varphi)\).
\end{definition}

\begin{definition}\label{delta-and-lambda-f}
	Let \(\langle \Sigma, \langle M, \otimes, \bar{1} \rangle, Q, s, F, \delta, \lambda, \iota, \rho, f, \varphi \rangle\) be a failure transducer. We define the \emph{completed transition function} \(\delta_f \colon Q \times \Sigma \to Q\) and the \emph{completed output function} \(\lambda_f \colon Q \times \Sigma \to M\) as the smallest with respect to inclusion functions \(\delta' \colon Q \times \Sigma \to Q\) and \(\lambda' \colon Q \times \Sigma \to M\) such that
	\begin{align*}
		\delta'(q, \sigma) &\coloneqq
		\begin{cases}
	  		\delta(q, \sigma) & \text{if } \defined\delta(q, \sigma) \\
			\delta'(f(q), \sigma) & \text{otherwise}
	  	\end{cases}
		\\
		\lambda'(q, \sigma) &\coloneqq
          	\begin{cases}
	  		\lambda(q, \sigma) & \text{if } \defined\lambda(q, \sigma) \\
			\varphi(q) \otimes \lambda'(f(q), \sigma) & \text{otherwise}
	  	\end{cases}
	\end{align*}
\end{definition}

\begin{definition}
	The \emph{expanded transducer} of a failure transducer \(\mathcal{F} \coloneqq \langle \Sigma, \mathcal{M},\allowbreak Q, s, F, \delta, \lambda, \iota, \rho, f, \varphi \rangle\) is the transducer \(\mathcal{T} \coloneqq \langle \Sigma, \mathcal{M}, Q, s, F, \delta_f, \lambda_f, \iota, \rho \rangle\). For each \(q \in Q\) we define \(\mathcal{O}_\mathcal{F}^q \coloneqq \mathcal{O}_\mathcal{T}^q\). The function \(\mathcal{O}_\mathcal{F} \coloneqq \mathcal{O}_\mathcal{T}\) is called \emph{the function represented by the failure transducer \(\mathcal{F}\)}. A state \(q \in Q\) is \emph{co-accessible} in \(\mathcal{F}\) if it is co-accessible in \(\mathcal{T}\).
\end{definition}

We call a failure transducer \emph{probabilistic} (\emph{stochastic}, \emph{canonical} with respect to a semiring \(\mathcal{S}\)) if its corresponding expanded transducer is probabilistic (stochastic, canonical with respect to a semiring \(\mathcal{S}\)). Note that, unlike stochastic transducers, stochastic failure transducers may have failure outputs greater than \(1\).

\begin{definition}
	Let \(\langle \Sigma, \mathcal{M}, Q, s, F, \delta, \lambda, \iota, \rho, f, \varphi \rangle\) be a failure transducer without failure cycles. We define the function \(\level_f \colon Q \to \mathbb{N}\) for \(q \in Q\) as
	\[\level_f(q) \coloneqq
	  \begin{cases}
	  	0 & \text{if } \neg\defined f(q) \\
		\level_f(f(q)) + 1 & \text{otherwise}
	  \end{cases}\]
\end{definition}

\begin{definition}
	A failure transducer \(\langle \Sigma, \mathcal{M}, Q, s, F, \delta, \lambda, \iota, \rho, f, \varphi \rangle\) is \emph{monotonic} if for every \(q \in \Dom(f)\)
	\begin{itemize}
		\item \(q \in F \implies f(q) \in F\);
		\item \((\forall a \in \Sigma)(\defined \delta(q, a) \implies \defined \delta(f(q), a))\).
	\end{itemize}
\end{definition}

\begin{proposition}
	Let \(\mathcal{F} \coloneqq \langle \Sigma, \langle M, \otimes, \bar{1} \rangle, Q, s, F, \delta, \lambda, \iota, \rho, f, \varphi \rangle\) be a monotonic failure transducer. Let \(C \coloneqq \{q \in Q \mid (\exists n \in \mathbb{N}^+)(f^{(n)}(q) = q)\}\), where \(f^{(n)}\) is \(f\) composed with itself \(n\) times. Then the failure transducer \(\mathcal{F'} \coloneqq \langle \Sigma, \mathcal{M}, Q, s, F, \delta,\allowbreak \lambda, \iota, \rho, f', \varphi' \rangle\), where \(f' \coloneqq f\restriction_{Q \setminus C}\) and \(\varphi' \coloneqq \varphi\restriction_{Q \setminus C}\), is such that \(\mathcal{O}_\mathcal{F'} = \mathcal{O}_\mathcal{F}\).
	\begin{proof}
		It is enough to show that for every \(q \in Q\) and \(\sigma \in \Sigma\)
		\begin{align*}
			\defined\delta_f(q, \sigma) &\iff \defined\delta_{f'}(q, \sigma); \\
			\defined\delta_f(q, \sigma) &\implies \delta_f(q, \sigma) = \delta_{f'}(q, \sigma) \land \lambda_f(q, \sigma) = \lambda_{f'}(q, \sigma).
		\end{align*}
		Since \(\mathcal{F'}\) has no failure cycles, we will proceed by induction on \(\level_{f'}(q)\).
		
		First, suppose \(\level_{f'}(q) = 0\), i.e. \(q \notin \Dom(f')\). If \(\defined\delta_{f'}(q, \sigma)\), then \(\defined\delta(q, \sigma)\) and therefore \(\defined\delta_f(q, \sigma)\). Suppose \(\defined\delta_f(q, \sigma)\). If \(\defined\delta(q, \sigma)\), then obviously \(\delta_{f'}(q, \sigma)\) is defined, \(\delta_f(q, \sigma) = \delta(q, \sigma) = \delta_{f'}(q, \sigma)\) and \(\lambda_f(q, \sigma) = \lambda(q, \sigma) = \lambda_{f'}(q, \sigma)\). If \(\neg\defined\delta(q, \sigma)\), then \(q \in \Dom(f)\setminus \Dom(f')\), therefore \(q \in C\) and \((\exists n \in \mathbb{N}^+)(f^{(n)}(q) = q)\). Since \(\mathcal{F}\) is monotonic, \((\forall 1 \le i \le n)(\neg\defined\delta(f^{(i)}(q), \sigma))\). By induction on \(i\) it is easily seen that \((\forall 1 \le i \le n)(\delta_f(q, \sigma) = \delta_f(f^{(i)}(q), \sigma))\). This contradicts with the minimality of \(\delta_f\) with respect to inclusion, because the function \(\delta' \coloneqq \delta_f\restriction_{\Dom(\delta_f) \setminus \{\langle q, \sigma \rangle\}}\) satisfies the conditions of \autoref{delta-and-lambda-f} and is strictly included in \(\delta_f\). Therefore, \(\defined\delta_f(q, \sigma)\) implies \(\defined\delta(q, \sigma)\).

		Now, suppose \(\level_{f'}(q) > 0\), i.e. \(q \in \Dom(f')\). By definition we have that \(f'(q) = f(q)\) and \(\varphi'(q) = \varphi(q)\). If \(\defined\delta(q, \sigma)\), then the reasoning is the same as in the base case. Suppose \(\neg\defined\delta(q, \sigma)\). Then using the inductive hypothesis we obtain
		\begin{align*}
			\defined\delta_f(q, \sigma) &\iff \defined\delta_f(f(q), \sigma) \iff \defined\delta_{f'}(f(q), \sigma) \iff \defined\delta_{f'}(f'(q), \sigma) \\
			&\iff \defined\delta_{f'}(q, \sigma), \\
			\delta_f(q, \sigma) &= \delta_f(f(q), \sigma) = \delta_{f'}(f(q), \sigma) = \delta_{f'}(f'(q), \sigma) = \delta_{f'}(q, \sigma), \\
			\lambda_f(q, \sigma) &= \varphi(q) \otimes \lambda_f(f(q), \sigma) = \varphi(q) \otimes \lambda_{f'}(f(q), \sigma) = \varphi'(q) \otimes \lambda_{f'}(f'(q), \sigma) \\
			&= \lambda_{f'}(q, \sigma).
		\end{align*}
	\end{proof}
\end{proposition}

In what follows, we will assume that every monotonic failure transducer that we consider has no failure cycles since they can be efficiently removed.

\section{Composition of Conditional Probabilistic Transducers with Probabilistic Failure Transducers}

Let \(\mathcal{M} \coloneqq \langle M, \otimes, \bar{1} \rangle\) be a commutative monoid, \(\mathcal{T} \coloneqq \langle \Sigma, \Omega^* \times \mathcal{M}, Q_1, s_1, F_1, \delta_1,\allowbreak \lambda_1, \iota_1, \rho_1 \rangle\) be a transducer and \(\mathcal{F} \coloneqq \langle \Omega, \mathcal{M}, Q_2, s_2, F_2, \delta_2, \lambda_2, \iota_2, \rho_2, f_2, \varphi_2 \rangle\) be a failure transducer without failure cycles. The following construction is an extension of the composition of transducers \cite{Mohri2008} and the intersection of weighted finite automata with failure transitions \cite{AllauzenFailure}.

\begin{definition}\label{generic-composition}
	The \emph{composition} of \(\mathcal{T}\) and \(\mathcal{F}\) is the failure transducer \(\mathcal{T} \circ \mathcal{F} \coloneqq \langle \Sigma, \mathcal{M}, Q_1 \times Q_2, s, F, \delta, \lambda, \iota, \rho, f,\allowbreak \varphi \rangle\), where
	\begin{align*}
		s \coloneqq {}& \langle s_1, {\delta_2}_{f_2}^{*}(s_2, \Proj_1(\iota_1)) \rangle, \\
		F \coloneqq {}& \{\langle p_1, p_2 \rangle \mid p_1 \in F_1, \exists q_2 \in F_2: \langle p_2, \Proj_1(\rho_1(p_1)), q_2 \rangle \in {\delta_2}_{f_2}^{*}\}, \\
		\delta \coloneqq {}& \{\langle \langle p_1, p_2 \rangle, a, \langle q_1, p_2 \rangle \rangle \mid \langle p_1, a, q_1 \rangle \in \delta_1, \exists o_1 \in M: \langle p_1, a, \langle \varepsilon, o_1 \rangle \rangle \in \lambda_1, p_2 \in Q_2\} \cup \\
		&\{\langle \langle p_1, p_2 \rangle, a, \langle q_1, q_2 \rangle \rangle \mid \langle p_1, a, q_1 \rangle \in \delta_1, \exists \omega \in \Omega, \exists \alpha \in \Omega^*, \exists o_1 \in M:\\
		& \hspace{3.37cm}\langle p_1, a, \langle \omega\alpha, o_1 \rangle \rangle \in \lambda_1 \land \defined\delta_2(p_2, \omega) \land \langle p_2, \omega\alpha, q_2 \rangle \in {\delta_2}_{f_2}^{*}\}, \\
		\lambda \coloneqq {}& \{\langle \langle p_1, p_2 \rangle, a, o_1 \rangle \mid \langle p_1, a, \langle \varepsilon, o_1 \rangle \rangle \in \lambda_1, p_2 \in Q_2\} \cup \\
		&\{\langle \langle p_1, p_2 \rangle, a, o_1 \otimes o_2 \rangle \mid \exists \omega \in \Omega, \exists \alpha \in \Omega^*: \langle p_1, a, \langle \omega\alpha, o_1 \rangle \rangle \in \lambda_1 \land \defined\delta_2(p_2, \omega) \land \\
		&\hspace{3.45cm}\langle p_2, \omega\alpha, o_2 \rangle \in {\lambda_2}_{f_2}^{*}\}, \\
		\iota \coloneqq {}& \Proj_2(\iota_1) \otimes \iota_2 \otimes {\lambda_2}_{f_2}^{*}(s_2, \Proj_1(\iota_1)), \\
		\rho \coloneqq {}& \{\langle \langle p_1, p_2 \rangle, o_1 \otimes o_2 \otimes o_3 \rangle \mid \langle p_1, p_2 \rangle \in F, \exists \alpha \in \Omega^*:  \langle p_1, \langle \alpha, o_1 \rangle \rangle \in \rho_1 \land \\ 
		&\hspace{3.85cm} \langle p_2, \alpha, o_2 \rangle \in {\lambda_2}_{f_2}^{*} \land \langle {\delta_2}_{f_2}^{*}(p_2, \alpha), o_3 \rangle \in \rho_2\}, \\
		f \coloneqq {}& \{\langle \langle p_1, p_2 \rangle, \langle p_1, q_2 \rangle \rangle \mid p_1 \in Q_1, \langle p_2, q_2 \rangle \in f_2\}, \\
		\varphi \coloneqq {}& \{\langle \langle p_1, p_2 \rangle, o_2 \rangle \mid p_1 \in Q_1, \langle p_2, o_2 \rangle \in \varphi_2\}.
	\end{align*}
\end{definition}

\begin{remark}
	\autoref{generic-composition} implies that \(\level_f(\langle p_1, p_2 \rangle) = \level_{f_2}(p_2)\) for every \(\langle p_1, p_2 \rangle \in Q_1 \times Q_2\), and since \(\mathcal{F}\) has no failure cycles, the resulting failure transducer \(\mathcal{T} \circ \mathcal{F}\) will not have failure cycles either.
\end{remark}

We proceed by giving a detailed correctness proof of the construction from \autoref{generic-composition}.

\begin{proposition}\label{generic-composition-delta-f}
	Let \(\langle p_1, p_2 \rangle \in Q_1 \times Q_2\) and \(a \in \Sigma\). Then
	\begin{align*}
		\defined\delta_f(\langle p_1, p_2 \rangle, a) \implies (&\delta_f(\langle p_1, p_2 \rangle, a) = \langle \delta_1(p_1, a), {\delta_2}_{f_2}^*(p_2, \beta) \rangle \land \\
		&\lambda_f(\langle p_1, p_2 \rangle, a) = o_1 \otimes {\lambda_2}_{f_2}^*(p_2, \beta)),
	\end{align*}
	where \(\langle \beta, o_1 \rangle \coloneqq \lambda_1(p_1, a)\).
	\begin{proof}
		We prove it by induction on \(\level_f(\langle p_1, p_2 \rangle)\).
		
		First, suppose \(\level_f(\langle p_1, p_2 \rangle) = 0\). Then \(\delta_f(\langle p_1, p_2 \rangle, a) = \delta(\langle p_1, p_2 \rangle, a)\) and \(\lambda_f(\langle p_1, p_2 \rangle, a) = \lambda(\langle p_1, p_2 \rangle, a)\).
		\begin{enumerate}
			\item Suppose \(\beta = \varepsilon\). Then \(\delta(\langle p_1, p_2 \rangle, a) = \langle \delta_1(p_1, a), p_2 \rangle = \langle \delta_1(p_1, a), {\delta_2}_{f_2}^*(p_2, \varepsilon) \rangle\) and \(\lambda(\langle p_1, p_2 \rangle, a) = o_1 = o_1 \otimes {\lambda_2}_{f_2}^*(p_2, \varepsilon)\).
			\item Suppose \(\beta = \omega\alpha\), where \(\omega \in \Omega\). Then \(\delta(\langle p_1, p_2 \rangle, a) = \langle \delta_1(p_1, a), {\delta_2}_{f_2}^*(p_2, \omega\alpha) \rangle\) and \(\lambda(\langle p_1, p_2 \rangle, a) = o_1 \otimes {\lambda_2}_{f_2}^*(p_2, \omega\alpha)\).
		\end{enumerate}
		
		Now, suppose \(\level_f(\langle p_1, p_2 \rangle) > 0\). If \(\defined\delta(\langle p_1, p_2 \rangle, a)\), then the reasoning is the same as in the base case. Suppose \(\neg\defined\delta(\langle p_1, p_2 \rangle, a)\), i.e. \(\beta = \omega\alpha\), \(\omega \in \Omega\) and
		\begin{align*}
			&\delta_f(\langle p_1, p_2 \rangle, a) = \delta_f(f(\langle p_1, p_2 \rangle), a) = \delta_f(\langle p_1, f_2(p_2) \rangle, a), \\
			&\lambda_f(\langle p_1, p_2 \rangle, a) = \varphi(\langle p_1, p_2 \rangle) \otimes \lambda_f(f(\langle p_1, p_2 \rangle), a) = \varphi_2(p_2) \otimes \lambda_f(\langle p_1, f_2(p_2) \rangle, a).
		\end{align*}
		Since \(\level_f(\langle p_1, f_2(p_2) \rangle) < \level_f(\langle p_1, p_2 \rangle)\), the inductive hypothesis holds for \(\langle p_1, f_2(p_2) \rangle\), i.e.
		\begin{align*}
			&\delta_f(\langle p_1, f_2(p_2) \rangle, a) = \langle \delta_1(p_1, a), {\delta_2}_{f_2}^*(f_2(p_2), \omega\alpha) \rangle, \\
			&\lambda_f(\langle p_1, f_2(p_2) \rangle, a) = o_1 \otimes {\lambda_2}_{f_2}^*(f_2(p_2), \omega\alpha).
		\end{align*}
		We also have that \(\neg\defined\delta_2(p_2, \omega)\), because otherwise \(\delta(\langle p_1, p_2 \rangle, a)\) would be defined. Therefore,
		\begin{align*}
			\delta_f(\langle p_1, p_2 \rangle, a) &= \delta_f(\langle p_1, f_2(p_2) \rangle, a) \\
			&= \langle \delta_1(p_1, a), {\delta_2}_{f_2}^*(f_2(p_2), \omega\alpha) \rangle \\
			&= \langle \delta_1(p_1, a), {\delta_2}_{f_2}^*(p_2, \omega\alpha) \rangle, \\
			\lambda_f(\langle p_1, p_2 \rangle, a) &= \varphi_2(p_2) \otimes \lambda_f(\langle p_1, f_2(p_2) \rangle, a) \\
			&= \varphi_2(p_2) \otimes o_1 \otimes {\lambda_2}_{f_2}^*(f_2(p_2), \omega\alpha) \\
			&= o_1 \otimes {\lambda_2}_{f_2}^*(p_2, \omega\alpha).
		\end{align*}
	\end{proof}
\end{proposition}

\begin{proposition}\label{generic-composition-delta-f-star}
	Let \(\langle p_1, p_2 \rangle \in Q_1 \times Q_2\) and \(\alpha \in \Sigma^*\). Then
	\begin{align*}
		\defined\delta_f^*(\langle p_1, p_2 \rangle, \alpha) \implies (&\delta_f^*(\langle p_1, p_2 \rangle, \alpha) = \langle \delta_1^*(p_1, \alpha), {\delta_2}_{f_2}^*(p_2, \beta) \rangle \land \\
		&\lambda_f^*(\langle p_1, p_2 \rangle, \alpha) = o_1 \otimes {\lambda_2}_{f_2}^*(p_2, \beta)),
	\end{align*}
	where \(\langle \beta, o_1 \rangle \coloneqq \lambda_1^*(p_1, \alpha)\).
	\begin{proof}
		We prove it by induction on \(|\alpha|\).
		
		First, suppose \(\alpha = \varepsilon\). Then \(\langle \beta, o_1 \rangle = \langle \varepsilon, \bar{1} \rangle\) and
		\begin{align*}
			&\delta_f^*(\langle p_1, p_2 \rangle, \varepsilon) = \langle p_1, p_2 \rangle = \langle \delta_1^*(p_1, \varepsilon), {\delta_2}_{f_2}^*(p_2, \varepsilon) \rangle, \\
			&\lambda_f^*(\langle p_1, p_2 \rangle, \varepsilon) = \bar{1} = o_1 \otimes {\lambda_2}_{f_2}^*(p_2, \varepsilon).
		\end{align*}
		
		Now, suppose \(\alpha = \alpha' a\), \(a \in \Sigma\) and the statement is true for every word with length \(|\alpha'|\). Then
		\begin{align*}
			\delta_f^*(\langle p_1, p_2 \rangle, \alpha) &= \delta_f(\delta_f^*(\langle p_1, p_2 \rangle, \alpha'), a) \\
			&= \delta_f(\langle \delta_1^*(p_1, \alpha'), {\delta_2}_{f_2}^*(p_2, \beta') \rangle, a) \\
			&= \langle \delta_1(\delta_1^*(p_1, \alpha'), a), {\delta_2}_{f_2}^*({\delta_2}_{f_2}^*(p_2, \beta'), \beta'') \rangle \\
			&= \langle \delta_1^*(p_1, \alpha), {\delta_2}_{f_2}^*(p_2, \beta) \rangle, \\
			\lambda_f^*(\langle p_1, p_2 \rangle, \alpha) &= \lambda_f^*(\langle p_1, p_2 \rangle, \alpha') \otimes \lambda_f(\delta_f^*(\langle p_1, p_2 \rangle, \alpha'), a) \\
			&= o_1' \otimes {\lambda_2}_{f_2}^*(p_2, \beta') \otimes \lambda_f(\langle \delta_1^*(p_1, \alpha'), {\delta_2}_{f_2}^*(p_2, \beta') \rangle, a) \\
			&= o_1' \otimes {\lambda_2}_{f_2}^*(p_2, \beta') \otimes o_1'' \otimes {\lambda_2}_{f_2}^*({\delta_2}_{f_2}^*(p_2, \beta'), \beta'') \\
			&= o_1 \otimes {\lambda_2}_{f_2}^*(p_2, \beta),
		\end{align*}
		where \(\langle \beta', o_1' \rangle \coloneqq \lambda_1^*(p_1, \alpha')\), \(\langle \beta'', o_1'' \rangle \coloneqq \lambda_1(\delta_1^*(p_1, \alpha'), a)\) and
		\begin{align*}
			\langle \beta, o_1 \rangle \coloneqq \langle \beta'\beta'', o_1' o_1'' \rangle = \lambda_1^*(p_1, \alpha') \odot \lambda_1(\delta_1^*(p_1, \alpha'), a) = \lambda_1^*(p_1, \alpha).
		\end{align*}
	\end{proof}
\end{proposition}

\begin{proposition}\label{generic-composition-delta-f-rl}
	Let \(\langle p_1, p_2 \rangle \in Q_1 \times Q_2\) and \(a \in \Sigma\). Then
	\[\defined\delta_1(p_1, a) \land \defined{\delta_2}_{f_2}^*(p_2, \Proj_1(\lambda_1(p_1, a))) \implies \defined\delta_f(\langle p_1, p_2 \rangle, a).\]
	\begin{proof}
		Analogous to the proof of \autoref{generic-composition-delta-f} in the opposite direction.
	\end{proof}
\end{proposition}

\begin{proposition}\label{generic-composition-delta-f-star-rl}
	Let \(\langle p_1, p_2 \rangle \in Q_1 \times Q_2\) and \(\alpha \in \Sigma\). Then
	\[\defined\delta_1^*(p_1, \alpha) \land \defined{\delta_2}_{f_2}^*(p_2, \Proj_1(\lambda_1^*(p_1, \alpha))) \implies \defined\delta_f^*(\langle p_1, p_2 \rangle, \alpha).\]
	\begin{proof}
		Follows from \autoref{generic-composition-delta-f-rl} and is analogous to the proof of \autoref{generic-composition-delta-f-star} in the opposite direction.
	\end{proof}
\end{proposition}

\begin{proposition}\label{generic-composition-correctness-aux}
	Let \(p_1 \in Q_1\) and \(p_2 \in Q_2\). Then
	\begin{enumerate}
		\item \((\forall \alpha \in \Dom(\mathcal{O}_{\mathcal{T} \circ \mathcal{F}}^{\langle p_1, p_2 \rangle}))(\alpha \in \Dom(\mathcal{O}_\mathcal{T}^{p_1}) \land \Proj_1(\mathcal{O}_\mathcal{T}^{p_1}(\alpha)) \in \Dom(\mathcal{O}_\mathcal{F}^{p_2}))\);
		\item \((\forall \alpha \in \Dom(\mathcal{O}_\mathcal{T}^{p_1}))(\Proj_1(\mathcal{O}_\mathcal{T}^{p_1}(\alpha)) \in \Dom(\mathcal{O}_\mathcal{F}^{p_2}) \implies \alpha \in \Dom(\mathcal{O}_{\mathcal{T} \circ \mathcal{F}}^{\langle p_1, p_2 \rangle}))\);
		\item \((\forall \alpha \in \Dom(\mathcal{O}_{\mathcal{T} \circ \mathcal{F}}^{\langle p_1, p_2 \rangle}))(\mathcal{O}_{\mathcal{T} \circ \mathcal{F}}^{\langle p_1, p_2 \rangle}(\alpha) = o_1 \otimes \mathcal{O}_\mathcal{F}^{p_2}(\beta))\), where \(\langle \beta, o_1 \rangle \coloneqq \mathcal{O}_\mathcal{T}^{p_1}(\alpha)\);
		\item if \(\mathcal{T}\) is conditional probabilistic and \(\Proj_1(\Rng(\mathcal{O}_{\mathcal{T}}^{p_1})) \supseteq \Dom(\mathcal{O}_{\mathcal{F}}^{p_2})\) then \[\sum_{\alpha \in \Dom(\mathcal{O}_{\mathcal{T} \circ \mathcal{F}}^{\langle p_1, p_2 \rangle}) } \mathcal{O}_{\mathcal{T} \circ \mathcal{F}}^{\langle p_1, p_2 \rangle}(\alpha) = \sum_{\beta \in \Dom(\mathcal{O}_{\mathcal{F}}^{p_2})} \mathcal{O}_{\mathcal{F}}^{p_2}(\beta) \sum_{\alpha \in \Dom(\mathcal{O}_\mathcal{T}^{p_1}(\bullet \mid \beta))} \mathcal{O}_\mathcal{T}^{p_1}(\alpha \mid \beta).\]
	\end{enumerate}

	\begin{proof}
		\begin{enumerate}
			\item[]
			\item Let \(\alpha \in \Dom(\mathcal{O}_{\mathcal{T} \circ \mathcal{F}}^{\langle p_1, p_2 \rangle})\). Then there exists a final state \(\langle q_1, q_2 \rangle \in F\), such that \(\delta_f^*(\langle p_1, p_2 \rangle, \alpha) = \langle q_1, q_2 \rangle\). From \autoref{generic-composition-delta-f-star} it follows that \(\delta_1^*(p_1, \alpha) = q_1\) and \({\delta_2}_{f_2}^*(p_2, \Proj_1(\lambda_1^*(p_1, \alpha))) = q_2\). Since \(\langle q_1, q_2 \rangle \in F\), then \(q_1 \in F_1\) and \({\delta_2}_{f_2}^*(q_2, \Proj_1(\rho_1(q_1))) \in F_2\). In other words, \(\alpha \in \Dom(\mathcal{O}_\mathcal{T}^{p_1})\) and
			\[\Proj_1(\lambda_1^*(p_1, \alpha))\Proj_1(\rho_1(\delta_1^*(p_1, \alpha))) = \Proj_1(\mathcal{O}_\mathcal{T}^{p_1}(\alpha)) \in \Dom(\mathcal{O}_\mathcal{F}^{p_2}).\]
			\item Let \(\alpha \in \Dom(\mathcal{O}_\mathcal{T}^{p_1})\), \(\beta \coloneqq \Proj_1(\mathcal{O}_\mathcal{T}^{p_1}(\alpha))\) and \(\beta \in \Dom(\mathcal{O}_\mathcal{F}^{p_2})\). Then there exist \(\beta', \beta'' \in \Sigma^*\) and \(q_1 \in F_1\), such that \(\beta = \beta'\beta''\), \(\delta_1^*(p_1, \alpha) = q_1\), \(\lambda_1^*(p_1, \alpha) = \beta'\) and \(\rho_1(q_1) = \beta''\). Also, there exists \(q_2 \in Q_2\), such that \({\delta_2}_{f_2}^*(p_2, \beta') = q_2\) and \({\delta_2}_{f_2}^*(q_2, \beta'') \in F_2\). From \autoref{generic-composition-delta-f-star} and \autoref{generic-composition-delta-f-star-rl} it follows that \(\delta_f^*(\langle p_1, p_2 \rangle, \alpha) = \langle q_1, q_2 \rangle\). By definition \(\langle q_1, q_2 \rangle \in F\), therefore \(\alpha \in \Dom(\mathcal{O}_{\mathcal{T} \circ \mathcal{F}}^{\langle p_1, p_2 \rangle})\).
			\item Let \(\alpha \in \Dom(\mathcal{O}_{\mathcal{T} \circ \mathcal{F}}^{\langle p_1, p_2 \rangle})\). Then
			\begin{alignat*}{2}
				\mathcal{O}_{\mathcal{T} \circ \mathcal{F}}^{\langle p_1, p_2 \rangle}(\alpha) &={}&& \lambda_f^*(\langle p_1, p_2 \rangle, \alpha) \otimes \rho(\delta_f^*(\langle p_1, p_2 \rangle, \alpha)) \\
				&={}&& [o_1' \otimes {\lambda_2}_{f_2}^*(p_2, \beta')] \otimes \rho(\langle \delta_1^*(p_1, \alpha), {\delta_2}_{f_2}^*(p_2, \beta') \rangle) \\
				&={}&&  [o_1' \otimes {\lambda_2}_{f_2}^*(p_2, \beta')] \otimes \\
				&{}&&  [o_1'' \otimes {\lambda_2}_{f_2}^*({\delta_2}_{f_2}^*(p_2, \beta'), \beta'') \otimes \rho_2({\delta_2}_{f_2}^*({\delta_2}_{f_2}^*(p_2, \beta'), \beta'' ))] \\
				&={}&& [o_1' \otimes o_1''] \otimes [{\lambda_2}_{f_2}^*(p_2, \beta') \otimes {\lambda_2}_{f_2}^*({\delta_2}_{f_2}^*(p_2, \beta'), \beta'')] \otimes \\
				&{}&& \rho_2({\delta_2}_{f_2}^*({\delta_2}_{f_2}^*(p_2, \beta'), \beta'' )) \\
				&={}&& [o_1' \otimes o_1''] \otimes {\lambda_2}_{f_2}^*(p_2, \beta'\beta'') \otimes \rho_2({\delta_2}_{f_2}^*(p_2, \beta'\beta'')) \\
				&={}&& o_1 \otimes \mathcal{O}_{\mathcal{F}}^{p_2}(\beta),
			\end{alignat*}
			where \(\langle \beta', o_1' \rangle \coloneqq \lambda_1^*(p_1, \alpha)\), \(\langle \beta'', o_1'' \rangle \coloneqq \rho_1(\delta_1^*(p_1, \alpha))\) and
				\[\langle \beta, o_1 \rangle \coloneqq \langle \beta'\beta'', o_1' \otimes o_1'' \rangle = \lambda_1^*(p_1, \alpha) \odot \rho_1(\delta_1^*(p_1, \alpha)) =  \mathcal{O}^{p_1}_\mathcal{T}(\alpha).\]
			\item Let \(\mathcal{T}\) be conditional probabilistic and \(\Proj_1(\Rng(\mathcal{O}_{\mathcal{T}}^{p_1})) \supseteq \Dom(\mathcal{O}_{\mathcal{F}}^{p_2})\) . Then
			\begin{align*}
				&\sum_{\alpha \in \Dom(\mathcal{O}_{\mathcal{T} \circ \mathcal{F}}^{\langle p_1, p_2 \rangle} )} \mathcal{O}_{\mathcal{T} \circ \mathcal{F}}^{\langle p_1, p_2 \rangle}(\alpha) \\
				&= \sum_{\alpha \in \Dom(\mathcal{O}_{\mathcal{T} \circ \mathcal{F}}^{\langle p_1, p_2 \rangle})} \sum_{\beta \in \Proj_1(\mathcal{O}_\mathcal{T}^{p_1}(\alpha))} \mathcal{O}_\mathcal{T}^{p_1}(\alpha \mid \beta) \mathcal{O}_\mathcal{F}^{p_2}(\beta) \\
				&= \sum_{\beta \in \Dom(\mathcal{O}_\mathcal{F}^{p_2})} \sum_{\alpha \in \Dom(\mathcal{O}_\mathcal{T}^{p_1}(\bullet \mid \beta))} \mathcal{O}_\mathcal{T}^{p_1}(\alpha \mid \beta) \mathcal{O}_\mathcal{F}^{p_2}(\beta) \\
				&= \sum_{\beta \in \Dom(\mathcal{O}_\mathcal{F}^{p_2})} \mathcal{O}_\mathcal{F}^{p_2}(\beta) \sum_{\alpha \in \Dom(\mathcal{O}_\mathcal{T}^{p_1}(\bullet \mid \beta))} \mathcal{O}_\mathcal{T}^{p_1}(\alpha \mid \beta).
			\end{align*}
		\end{enumerate}
	\end{proof}
\end{proposition}

\begin{proposition}\label{generic-composition-correctness}
	\begin{enumerate}
		\item[]
		\item \((\forall \alpha \in \Dom(\mathcal{O}_{\mathcal{T} \circ \mathcal{F}}))(\alpha \in \Dom(\mathcal{O}_\mathcal{T}) \land \Proj_1(\mathcal{O}_\mathcal{T}(\alpha)) \in \Dom(\mathcal{O}_\mathcal{F}))\);
		\item \((\forall \alpha \in \Dom(\mathcal{O}_\mathcal{T}))(\Proj_1(\mathcal{O}_\mathcal{T}(\alpha)) \in \Dom(\mathcal{O}_\mathcal{F}) \implies \alpha \in \Dom(\mathcal{O}_{\mathcal{T} \circ \mathcal{F}}))\);
		\item \((\forall \alpha \in \Dom(\mathcal{O}_{\mathcal{T} \circ \mathcal{F}}))(\mathcal{O}_{\mathcal{T} \circ \mathcal{F}}(\alpha) = o_1 \otimes \mathcal{O}_\mathcal{F}(\beta))\), where \(\langle \beta, o_1 \rangle = \mathcal{O}_\mathcal{T}(\alpha)\);
		\item if \(\mathcal{T}\) is conditional probabilistic, \(\mathcal{F}\) is probabilistic and \(\Proj_1(\Rng(\mathcal{O}_\mathcal{T})) \supseteq \Dom(\mathcal{O}_\mathcal{F})\), then \(\mathcal{T} \circ \mathcal{F}\) is probabilistic;
		\item if \(\mathcal{F}\) is monotonic and for every \(p \in \Dom(f_2)\)
			\begin{align*}
				&(\forall \alpha \in \Proj_1(\Rng(\lambda_1)))(\defined{\delta_2}_{f_2}^*(p, \alpha) \implies \defined{\delta_2}_{f_2}^*(f_2(p), \alpha)), \\
				&(\forall \alpha \in \Proj_1(\Rng(\rho_1)))({\delta_2}_{f_2}^*(p, \alpha) \in F_2 \implies {\delta_2}_{f_2}^*(f_2(p), \alpha) \in F_2),
			\end{align*}
			then \(\mathcal{T} \circ \mathcal{F}\) is monotonic.
	\end{enumerate}
	\begin{proof}
		\begin{enumerate}
			\item[]
			\combine Follows from \autoref{generic-composition-correctness-aux} applied for the initial state \(\langle s_1, s_2 \rangle\).
			\item Let \(\alpha \in \Dom(\mathcal{O}_{\mathcal{T} \circ \mathcal{F}})\) and \(t_2 \coloneqq {\delta_2}_{f_2}^*(s_2, \Proj_1(\iota_1))\).Then
			\begin{alignat*}{2}
				\mathcal{O}_{\mathcal{T} \circ \mathcal{F}}(\alpha) &={}&& \iota \otimes \mathcal{O}_{\mathcal{T}\circ \mathcal{F}}^{\langle s_1, t_2 \rangle}(\alpha)\\
				& = {}&& [o_1' \otimes \iota_2 \otimes {\lambda_2}_{f_2}^*(s_2, \beta')] \otimes [o_1'' \otimes \mathcal{O}_{\mathcal{F}}^{t_2}(\beta'')] \\
				& = {}&& [o_1' \otimes o_1''] \otimes [\iota_2 \otimes {\lambda_2}_{f_2}^*(s_2, \beta') \otimes {\lambda_2}_{f_2}^*(t_2, \beta'') \otimes \rho_2({\delta_2}_{f_2}^*(t_2, \beta''))] \\
				& = {}&& o_1 \otimes \mathcal{O}_{\mathcal{F}}(\beta),
			\end{alignat*}
			where \(\langle \beta', o_1' \rangle \coloneqq \iota_1\), \(\langle \beta'', o_1'' \rangle \coloneqq \mathcal{O}_{\mathcal{T}}^{s_1}(\alpha)\) and
			\begin{align*}
				\langle \beta, o_1 \rangle \coloneqq \langle \beta'\beta'', o_1' \otimes o_1'' \rangle = \iota_1 \odot \lambda_1^*(s_1, \alpha) \odot \rho_1(\delta_1^*(s_1, \alpha)) =  \mathcal{O}_\mathcal{T}(\alpha).
			\end{align*}
			\item Let \(\mathcal{T}\) be conditional probabilistic, \(\mathcal{F}\) be probabilisitc and \(\Proj_1(\Rng(\mathcal{O}_\mathcal{T})) \supseteq \Dom(\mathcal{O}_\mathcal{F})\). Let \(\langle \beta', o_1 \rangle \coloneqq \iota_1\) and \(t_2 \coloneqq {\delta_2}_{f_2}^*(s_1, \beta')\). Then \(\beta'\) is a prefix of every word in \(\Dom(\mathcal{O}_\mathcal{F})\) and using \autoref{generic-composition-correctness-aux}
			\begin{align*}
				&\sum_{\alpha \in \Dom(\mathcal{O}_{\mathcal{T} \circ \mathcal{F}})} \mathcal{O}_{\mathcal{T} \circ \mathcal{F}}(\alpha) \\
				&= \iota \sum_{\alpha \in \Dom(\mathcal{O}_{\mathcal{T} \circ \mathcal{F}}^{\langle s_1, t_2 \rangle})} \mathcal{O}_{\mathcal{T} \circ \mathcal{F}}^{\langle s_1, t_2 \rangle}(\alpha) \\
				&= o_1 \iota_2 {\lambda_2}_{f_2}^{*}(s_2, \beta') \sum_{\beta'' \in \Dom(\mathcal{O}_\mathcal{F}^{t_2})} \mathcal{O}_\mathcal{F}^{t_2}(\beta'') \sum_{\alpha \in \Dom(\mathcal{O}_\mathcal{T}^{s_1}(\bullet \mid \beta''))} \mathcal{O}_\mathcal{T}^{s_1}(\alpha \mid \beta'') \\
				&= \sum_{\beta'' \in \Dom(\mathcal{O}_\mathcal{F}^{t_2})} \iota_2 {\lambda_2}_{f_2}^{*}(s_2, \beta') \mathcal{O}_\mathcal{F}^{t_2}(\beta'') \sum_{\alpha \in \Dom(\mathcal{O}_\mathcal{T}^{s_1}(\bullet \mid \beta''))} o_1 \mathcal{O}_\mathcal{T}^{s_1}(\alpha \mid \beta'') \\
				&= \sum_{\beta'' \in \Dom(\mathcal{O}_\mathcal{F}^{t_2})} \iota_2 \mathcal{O}_\mathcal{F}^{s_2}(\beta'\beta'') \sum_{\alpha \in \Dom(\mathcal{O}_\mathcal{T}^{s_1}(\bullet \mid \beta''))} o_1 \mathcal{O}_\mathcal{T}^{s_1}(\alpha \mid \beta'') \\
				&= \sum_{\beta \in \Dom(\mathcal{O}_\mathcal{F})} \mathcal{O}_\mathcal{F}(\beta) \sum_{\alpha \in \Dom(\mathcal{O}_\mathcal{T}(\bullet \mid \beta))} \mathcal{O}_\mathcal{T}(\alpha \mid \beta) \\
				& = \sum_{\beta \in \Dom(\mathcal{O}_\mathcal{F})} \mathcal{O}_\mathcal{F}(\beta) \\
				&= 1.
			\end{align*}
			\item Let \(\mathcal{F}\) be monotonic. Let \(\langle p_1, p_2 \rangle \in Q_1 \times Q_2\) and \(\langle r_1, r_2 \rangle \coloneqq f(\langle p_1, p_2 \rangle)\). Then by definition \(r_1 = p_1\) and \(r_2 = f_2(p_2)\). If \(\langle p_1, p_2 \rangle \in F\), then \(p_1 \in F_1\) and \({\delta_2}_{f_2}^*(p_2, \Proj_1(\rho_1(p_1))) \in F_2\). This means that \({\delta_2}_{f_2}^*(f_2(p_2), \Proj_1(\rho_1(p_1))) \in F_2\) and therefore \(\langle p_1, f_2(p_2) \rangle \in F\). Suppose that \(\langle \langle p_1, p_2 \rangle, a, \langle q_1, q_2 \rangle \rangle \in \delta\). If \(\Proj_1(\lambda_1(p_1, a)) = \varepsilon\), then \(\langle \langle p_1, f_2(p_2) \rangle, a, \langle q_1, f_2(p_2) \rangle \rangle \in \delta\). Let \(\omega \alpha \coloneqq \Proj_1(\lambda_1(p_1, a))\) and \(\omega \in \Omega\). Then \(\defined\delta_2(p_2, \omega)\) and therefore \(\defined\delta_2(f_2(p_2), \omega)\) because \(\mathcal{F}\) is monotonic. Also, \({\delta_2}_{f_2}^*(p_2, \omega\alpha)\) is defined, which implies that \(\defined{\delta_2}_{f_2}^*(f_2(p_2), \omega\alpha)\) and thus \(\delta(\langle p_1, f_2(p_2) \rangle, a)\) is defined.
		\end{enumerate}
	\end{proof}
\end{proposition}

\section{Iteration of Conditional Probabilistic Transducers}\label{kleene-star-section}

Let \(\mathcal{V} \coloneqq \langle \Sigma, \Omega^* \times \mathcal{M}, Q_{1}, s_{1}, F_{1}, \delta_{1}, \lambda_{1}, \iota_{1}, \rho_{1} \rangle\) be a transducer such that \\ \(\Proj_1(\Rng(\mathcal{O}_\mathcal{V})) = \Omega\) and \(\Dom(\mathcal{O}_\mathcal{V})\) is prefix-free\footnote{This condition can be easily satisfied by adding a new symbol \(\$ \not\in \Sigma\) to the alphabet, defining the final states to be \(\{q_\$\}\) with final output \(\langle \varepsilon, \bar{1} \rangle\), where \(q_\$ \not\in Q_1\), and replacing the final outputs of each state \(q_1 \in F_1\) with transitions to \(q_\$\) with input \(\$\) and output \(\langle \varepsilon, \rho(q_1) \rangle\).}. The fact that \(\Dom(\mathcal{O}_\mathcal{V})\) is prefix-free means that there are no transitions leaving a final state. We will assume that \(\mathcal{V}\) has the following additional properties:
\begin{enumerate}
	\item There are no transitions that enter \(s_1\), otherwise we can introduce a new super initial state with the same finality and outgoing transitions as \(s_1\).
	\item \(\varepsilon \not\in \Dom(\mathcal{O}_\mathcal{V})\) and \(s_1 \notin F_1\).
	\item \(\iota_1 = \langle \varepsilon, \bar{1} \rangle\), otherwise we can left multiply (with the monoid operation) the outputs of the transitions that leave the initial state.
	\item \(\Rng(\rho_1) = \{\langle \varepsilon, \bar{1} \rangle\}\), otherwise we can right multiply the outputs of the transitions that enter a final state.
\end{enumerate}

\begin{definition}\label{kleene-star}
	The \emph{Kleene-Star} of \(\mathcal{V}\) is the transducer \(\mathcal{V}^* \coloneqq \langle \Sigma, \Omega^* \times \mathcal{M}, Q_1 \setminus F_1,\allowbreak s_1, \{s_1\}, \delta_2, \lambda_1, \langle \varepsilon, \bar{1} \rangle, \{\langle s_1, \langle \varepsilon, \bar{1} \rangle \rangle \} \rangle\), where
	\begin{align*}
		\delta_2 \coloneqq {}&\delta_1\restriction_{(Q_1 \setminus F_1) \times \Sigma \times (Q_1 \setminus F_1)} \cup \{\langle p_1, a, s_1 \rangle \mid \langle p_1, a, q_1 \rangle \in \delta_1, q_1 \in F_1\}.
	\end{align*}
\end{definition}

\begin{proposition}\label{kleene-star-auxiliary}
	Let \(p \in Q_1 \setminus F_1\) and \(\langle p, \alpha, q \rangle \in \delta_1^*\). Then
	\begin{enumerate}
		\item \(q \in Q_1 \setminus F_1 \implies \delta_2^*(p, \alpha) = q\);
		\item \(q \in F_1 \implies \delta_2^*(p, \alpha) = s_1\).
	\end{enumerate}
	\begin{proof}
		We proceed by induction on \(|\alpha|\).
		
		First, suppose \(\alpha = \varepsilon\), i.e. \(q = p \in Q_1 \setminus F_1\). Then (1) \(\delta_2^*(p, \varepsilon) = p = q\) and (2) holds vacuously.
		
		Now, suppose \(\alpha = \alpha' a\), \(a \in \Sigma\). Let \(r \coloneqq \delta_1^*(p, \alpha')\) and \(q = \delta_1(r, a)\). Then \(r \in Q_1 \setminus F_1\), since final states have no outgoing transitions in \(\mathcal{V}\). By the inductive hypothesis \(\delta_2^*(p, \alpha') = r\).
		\begin{enumerate}
			\item Suppose \(q \in Q_1 \setminus F_1\). From \(\delta_2 \supseteq \delta_1\restriction_{(Q_1 \setminus F_1) \times \Sigma \times (Q_1 \setminus F_1)}\) it follows that \(\delta_2(r, a) = q\). Therefore, \(\delta_2^*(p, \alpha) = q\).
			\item Suppose \(q \in F_1\). Then from the definition of \(\delta_2\), \(\delta_2(r, a) = s_1\). Therefore, \(\delta_2^*(p, \alpha) = s_1\).
		\end{enumerate}
	\end{proof}
\end{proposition}

\begin{proposition}\label{kleene-star-auxiliary2}
	Let \(p \in Q_1 \setminus F_1\) and \(\langle p, \alpha, q \rangle \in \delta_2^*\) without going through \(s_1\) as an intermediate state. Then
	\begin{enumerate}
		\item \(q \in Q_1 \setminus F_1 \implies \delta_1^*(p, \alpha) = q\);
		\item \(q = s_1 \implies \delta_1^*(p, \alpha) \in F_1\).
	\end{enumerate}
	\begin{proof}
		Analogous to the proof of \autoref{kleene-star-auxiliary}  in the opposite direction.
	\end{proof}
\end{proposition}

\begin{proposition}\label{kleene-star-correctness}
	\begin{enumerate}
		\item[]
		\item \(\Dom(\mathcal{O}_\mathcal{V^*}) = \Dom(\mathcal{O}_\mathcal{V})^*\);
		\item \((\forall \alpha_1, \alpha_2, \ldots, \alpha_n \in \Dom(\mathcal{O}_\mathcal{V}))(\mathcal{O}_{\mathcal{V}^*}(\alpha_1 \alpha_2 \ldots \alpha_n) = \bigodot_{i = 1}^n \mathcal{O}_\mathcal{V}(\alpha_i))\);
		\item if \(\mathcal{V}\) is conditional probabilistic, then \(\mathcal{V}^*\) is also conditional probabilistic;
		\item if \(\mathcal{V}\) is conditional probabilistic and \(\mathcal{F}\) is a probabilistic failure transducer with alphabet \(\Omega\), then \(\mathcal{V}^* \circ \mathcal{F}\) is probabilistic;
		\item if \(\mathcal{F}\) is a monotonic failure transducer, then \(\mathcal{V}^* \circ \mathcal{F}\) is monotonic.
	\end{enumerate}
	\begin{proof}
		\begin{enumerate}
			\item[]
			\item Let \(\alpha \in \Dom(\mathcal{O}_{\mathcal{V}^*})\), i.e. \(\langle s_1, \alpha, s_1 \rangle \in \delta_2^*\). Then there exist \(\alpha_1, \alpha_2, \ldots, \alpha_n \in \Sigma^*\) such that \((\forall 1 \le i \le n)(\langle s_1, \alpha_i, s_1 \rangle \in \delta_2^*)\) without going through \(s_1\) as an intermediate state and \(\alpha = \alpha_1\alpha_2 \ldots \alpha_n\). Then using \autoref{kleene-star-auxiliary2} we obtain that \((\forall 1 \le i \le n)(\delta_1^*(s_1, \alpha_i) \in F_1)\). In other words, \((\forall 1 \le i \le n)(\alpha_i \in \Dom(\mathcal{O}_\mathcal{V}))\) and \(\alpha \in  \Dom(\mathcal{O}_\mathcal{V})^*\).
			
			Let \(\alpha_1, \alpha_2, \ldots, \alpha_n \in \Dom(\mathcal{O}_\mathcal{V})\), i.e. \(\alpha_1\alpha_2 \ldots \alpha_n \in \Dom(\mathcal{O}_\mathcal{V})^*\). Then for \(1 \le i \le n\) we have that \(\delta_1^*(s_1, \alpha_i) \in F_1\). The initial state \(s_1\) is non-final in \(\mathcal{V}\). Then by \autoref{kleene-star-auxiliary}, \(\delta_2^*(s_1, \alpha_i) = s_1\) for \(1 \le i \le n\). This implies that \(\delta_2^*(s_1, \alpha_1\alpha_2 \ldots \alpha_n) = s_1\), i.e. \(\alpha_1\alpha_2 \ldots \alpha_n \in \Dom(\mathcal{O}_{\mathcal{V}^*})\).
			
			\item Let \(\alpha_1, \alpha_2, \ldots, \alpha_n \in \Dom(\mathcal{O}_\mathcal{V})\). From \autoref{kleene-star-auxiliary} we have that \((\forall 1 \le i \le n)\allowbreak(\delta_2^*(s_1, \alpha_i) = s_1)\), which means that \((\forall 1 \le i \le n)(\delta_2^*(s_1, \alpha_1\alpha_2 \ldots \alpha_i) = s_1)\) and
			\begin{align*}
				\mathcal{O}_{\mathcal{V}^*}(\alpha_1\alpha_2 \ldots \alpha_n) &= \lambda_1^*(s_1, \alpha_1\alpha_2 \ldots \alpha_n) \\
				&= \bigodot_{i=1}^n \lambda_1^*(\delta_2^*(s_1, \alpha_1\alpha_2 \ldots \alpha_{i-1}), \alpha_i) \\
				&= \bigodot_{i=1}^n \lambda_1^*(s_1, \alpha_i) \\
				&= \bigodot_{i=1}^n \mathcal{O}_\mathcal{V}(\alpha_i).
			\end{align*}
			
			\item Let \(\beta \in \Proj_1(\Rng(\mathcal{O}_{\mathcal{V}^*}))\) and \(\beta = b_1 b_2 \ldots b_n\), where \(b_1, b_2, \ldots, b_n \in \Omega\). Then
			\begin{align*}
				&\sum_{\alpha \in \Dom(\mathcal{O}_{\mathcal{V}^*}(\bullet \mid \beta))} \mathcal{O}_{\mathcal{V}^*}(\alpha \mid \beta) \\
				&= \sum_{\alpha_1 \in \Dom(\mathcal{O}_\mathcal{V}(\bullet \mid b_1))} \cdots \sum_{\alpha_n \in \Dom(\mathcal{O}_\mathcal{V}(\bullet \mid b_n))} \mathcal{O}_{\mathcal{V}^*}(\alpha_1\alpha_2 \ldots \alpha_n \mid \beta) \\
				&= \sum_{\alpha_1 \in \Dom(\mathcal{O}_\mathcal{V}(\bullet \mid b_1))} \cdots \sum_{\alpha_n \in \Dom(\mathcal{O}_\mathcal{V}(\bullet \mid b_n))} \prod_{i=1}^n \mathcal{O}_\mathcal{V}(\alpha_i \mid \beta_i) \\
				&= \sum_{\alpha_1 \in \Dom(\mathcal{O}_\mathcal{V}(\bullet \mid b_1))} \mathcal{O}_\mathcal{V}(\alpha_1 \mid \beta_1) \cdots \sum_{\alpha_n \in \Dom(\mathcal{O}_\mathcal{V}(\bullet \mid b_n))} \mathcal{O}_\mathcal{V}(\alpha_n \mid \beta_n) \\
				&= 1.
			\end{align*}
			\item By (3), \(\mathcal{V}^*\) is conditional probabilistic. \(\Proj_1(\Rng(\mathcal{O}_\mathcal{V})) = \Omega\) and
			\[\Proj_1(\Rng(\mathcal{O}_{\mathcal{V}^*})) = \Proj_1(\Rng(\mathcal{O}_\mathcal{V}))^* = \Omega^* \supseteq \Dom(\mathcal{O}_\mathcal{F}).\]
			From \autoref{generic-composition-correctness} it follows that \(\mathcal{V}^* \circ \mathcal{F}\) is probabilistic.
			\item Let \(\rho_2\) be the final output function of \(\mathcal{V}^*\). By definition \(\Proj_1(\Rng(\lambda_1)) \subseteq \Omega \cup \{\varepsilon\}\) and \(\Proj_1(\Rng(\rho_2)) = \{\varepsilon\}\). Let \(p \in \Dom(f_2)\).
			
			Let \(\alpha \in \Proj_1(\Rng(\lambda_1))\) and \(\defined{\delta_2}_{f_2}^*(p, \alpha)\). If \(\alpha = \varepsilon\), then \({\delta_2}_{f_2}^*(f_2(p), \alpha)\) is also defined. Suppose \(\alpha \in \Omega\). Then \({\delta_2}_{f_2}^*(p, \alpha) = {\delta_2}_{f_2}(p, \alpha)\). If \(\delta_2(p, \alpha)\) is defined, then by the monotonicity of \(\mathcal{F}\), \(\delta_2(f_2(p), \alpha)\) is defined and therefore \(\defined{\delta_2}_{f_2}^*(f_2(p), \alpha)\). Otherwise, if \(\neg\defined\delta_2(p, \alpha)\), then \({\delta_2}_{f_2}(p, \alpha) = {\delta_2}_{f_2}(f_2(p), \alpha)\) by definition. Thus, \({\delta_2}_{f_2}^*(f_2(p), \alpha)\) is defined.
			
			Let \(\alpha \in \Proj_1(\Rng(\rho_1))\) and \({\delta_2}_{f_2}^*(p, \alpha) = {\delta_2}_{f_2}^*(p, \varepsilon) = p \in F_2\). From the monotonicity of \(\mathcal{F}\) it follows that
			\[{\delta_2}_{f_2}^*(f_2(p), \alpha) = {\delta_2}_{f_2}^*(f_2(p), \varepsilon) = f_2(p) \in F_2.\]
			
			Now, \autoref{generic-composition-correctness} implies that \(\mathcal{V}^* \circ \mathcal{F}\) is monotonic.
		\end{enumerate}
	\end{proof}
\end{proposition}

\section{Specialized Composition}\label{special-composition}

Let \(\mathcal{V} \coloneqq \langle \Sigma, \Omega^* \times \mathcal{R}, Q_{1}, s_{1}, F_{1}, \delta_{1}, \lambda_{1}, \iota_{1}, \rho_{1} \rangle\) be a trim (i.e. \((\forall q \in Q_{1})(\exists \alpha, \beta \in \Sigma^*)(\delta_{1}^*(s_1, \alpha) = q \land \delta_1^*(q, \beta) \in F_{1})\)) conditional probabilistic transducer with the properties from \autoref{kleene-star-section}. Let  \(\mathcal{F} \coloneqq \langle \Omega, \mathcal{R}, Q_{2}, s_{2}, F_{2}, \delta_{2}, \lambda_{2}, \iota_{2}, \rho_{2},\allowbreak f_{2}, \varphi_{2} \rangle\) be a monotonic probabilistic failure transducer in which every state is co-accessible and \(\mathcal{V}^* \coloneqq \langle \Sigma, \Omega^* \times \mathcal{R}, Q_{3}, s_{3},\allowbreak F_{3}, \delta_{3}, \lambda_{3}, \iota_{3}, \rho_{3} \rangle\) be the Kleene-Star of \(\mathcal{V}\) from \autoref{kleene-star}. Let \(\mathcal{V}^* \circ \mathcal{F} \coloneqq \langle \Sigma, \mathcal{R}, Q_4, s_4, F_4, \delta_4, \lambda_4, \iota_4,\allowbreak \rho_4, f_4, \varphi_4 \rangle\) be the composition of \(\mathcal{V}^*\) and \(\mathcal{F}\) from \autoref{generic-composition}. In this section a more efficient construction for the composition \(\mathcal{V}^*\circ \mathcal{F}\) will be shown in which the creation of non-co-accessible states is avoided. 

\begin{definition}
	Let \(w \in \Omega\). We define
	\begin{align*}
		\Delta_\omega & \coloneqq \{\langle p_1, a, q_1 \rangle \in \delta_1 \mid \Proj_1(\lambda_1(p_1, a)) = \omega\}, \\ 
		Q_w^l & \coloneqq \bigcup_{\langle p_1, a, q_1 \rangle \in \Delta_\omega} \{l_1 \mid (\exists \alpha \in \Sigma^\ast)(\langle l_1, \alpha, p_1 \rangle \in \delta_1^*)\}, \\
		Q_w^r & \coloneqq \bigcup_{\langle p_1, a, q_1 \rangle \in \Delta_\omega} \{r_1 \mid (\exists \alpha \in \Sigma^\ast)(\langle q_1, \alpha, r_1 \rangle \in \delta_1^*)\}.
	\end{align*}
\end{definition}

\begin{proposition}\label{left-right-property}
	\begin{enumerate}
		\item[]
		\item \(\bigcup_{\omega \in \Omega} Q_\omega^l \cup Q_\omega^r = Q_1\);
		\item \((\forall \omega \in \Omega)(\forall \omega' \in \Omega)(Q_\omega^l \cap Q_{\omega'}^r = \varnothing)\);
		\item \((\forall \omega \in \Omega)(\forall p \in Q_\omega^\xi)(\forall a \in \Sigma)(\Proj_1(\lambda_1(p, a)) = \varepsilon \implies \delta_1(p, a) \in Q_\omega^\xi)\),\\ where \(\xi \in \{l, r\}\);
		\item 
		\((\forall \omega \in \Omega)(\forall p, q \in Q_\omega^\xi)((\exists a \in \Sigma)(\langle p, a, q \rangle \in \delta_1) \implies \Proj_1(\lambda_1(p, a)) = \varepsilon)\), where \(\xi \in \{l, r\}\).
	\end{enumerate}
\end{proposition}

\begin{definition}
	We define \(E \colon Q_1 \to Q_1\) for \(p \in Q_1\) as
	\[E(p) \coloneqq
	  \begin{cases}
	  	s_1 & \text{if } p \in F_1 \\
	  	p & \text{otherwise}
	  \end{cases}\]
\end{definition}

\begin{definition}\label{optimized-composition}
	We define \(\mathcal{W} \coloneqq \langle \Sigma, \mathcal{R}, Q, \langle s_1, s_2 \rangle, \{s_1\} \times F_2, \delta, \lambda, \iota_2, \rho, f, \varphi \rangle\), where
	\begin{align*}
		Q \coloneqq {}& \bigcup_{\langle p_2, \omega, q_2 \rangle \in \delta_2} Q_\omega^l \times \{p_2\} \cup E(Q_\omega^r) \times \{q_2\}, \\
		\delta \coloneqq {}& \bigcup_{\langle p_2, \omega, q_2 \rangle \in \delta_2} 
		\begin{aligned}[t]
			&\{\langle \langle p_1, p_2 \rangle, a, \langle q_1, p_2 \rangle \rangle \mid p_1, q_1 \in Q_\omega^l, \langle p_1, a, q_1 \rangle \in \delta_1\} \cup \\
			&\{\langle \langle p_1, p_2 \rangle, a, \langle E(q_1), q_2 \rangle \rangle \mid \langle p_1, a, q_1 \rangle \in \Delta_\omega\} \cup \\
			&\{\langle \langle p_1, q_2 \rangle, a, \langle E(q_1), q_2 \rangle \rangle \mid p_1, q_1 \in Q_\omega^r, \langle p_1, a, q_1 \rangle \in \delta_1\},
		\end{aligned}
		\\
		\lambda \coloneqq {}& \bigcup_{\langle p_2, \omega, q_2 \rangle \in \delta_2}
		\begin{aligned}[t]
			&\{\langle \langle p_1, p_2 \rangle, a, o_1 \rangle \mid p_1 \in Q_\omega^l, \langle p_1, a, \langle \varepsilon, o_1 \rangle \rangle \in \lambda_1\} \cup \\
			&\{\langle \langle p_1, p_2 \rangle, a, o_1o_2 \rangle \mid \langle p_1, a, \langle \omega, o_1 \rangle \rangle \in \lambda_1, \langle p_2, \omega, o_2 \rangle \in \lambda_2\} \cup \\
			&\{\langle \langle p_1, q_2 \rangle, a, o_1 \rangle \mid p_1 \in Q_\omega^r, \langle p_1, a, \langle \varepsilon, o_1 \rangle \rangle \in \lambda_1\},
		\end{aligned}
		\\
		\rho \coloneqq {}& \{\langle \langle s_1, p_2 \rangle, o_2 \rangle \mid \langle p_2, o_2 \rangle \in \rho_2\}, \\
		f \coloneqq {}& \bigcup_{\langle p_2, \omega, q_2 \rangle \in \delta_2} \{\langle \langle p_1, p_2 \rangle, \langle p_1, r_2 \rangle \rangle \mid p_1 \in Q_\omega^l, \langle p_2, r_2 \rangle \in f_2\}, \\
		\varphi \coloneqq {}& \bigcup_{\langle p_2, \omega, q_2 \rangle \in \delta_2} \{\langle \langle p_1, p_2 \rangle, o_2 \rangle \mid p_1 \in Q_\omega^l, \langle p_2, o_2 \rangle \in \varphi_2\}.
	\end{align*}
\end{definition}

\begin{remark}
	\autoref{optimized-composition} implies that \(\level_f(\langle p_1, p_2 \rangle) = \level_{f_2}(p_2)\) for every \(\langle p_1, p_2 \rangle \in Q_1 \times Q_2\), and since \(\mathcal{F}\) has no failure cycles, the resulting failure transducer \(\mathcal{W}\) will not have failure cycles either.
\end{remark}

\begin{proposition}\label{delta-4-implies-delta}
	Let \(p \in Q\) and \(a \in \Sigma\). Then
		\[\defined\delta_4(p, a) \implies \defined\delta(p, a).\]
	\begin{proof}
		Let \(\langle p_1, p_2 \rangle \coloneqq p\) and \(\defined\delta_4(\langle p_1, p_2 \rangle, a)\). We consider two cases for the transition according to \autoref{generic-composition}:
		\begin{enumerate}
			\item Suppose \(\langle p_1, a \rangle \in \Dom(\delta_3)\) and \(\Proj_1(\lambda_3(p_1, a)) = \Proj_1(\lambda_1(p_1, a)) = \varepsilon\). Then \(\langle p_1, a \rangle \in \Dom(\delta_1)\). \(p_1 \in Q_1\), therefore exists \(\omega \in \Omega\) such that \(p_1 \in Q_\omega^\xi\), \(\xi \in \{l, r\}\). From \autoref{left-right-property} it follows that \(\delta_1(p_1, a) \in Q_\omega^\xi\). Now, by definition \(\defined\delta(p, a)\).
			\item Suppose \(\langle p_1, a \rangle \in \Dom(\delta_3)\), \(\omega \coloneqq \Proj_1(\lambda_3(p_1, a)) = \Proj_1(\lambda_1(p_1, a))\), \(\omega \in \Omega\) and \(\defined\delta_2(p_2, \omega)\). Then \((\exists q_1 \in Q_1)(\langle p_1, a, q_1 \rangle \in \Delta_\omega)\), which implies that \(\defined\delta(p, a)\).
		\end{enumerate}
	\end{proof}
\end{proposition}

\begin{proposition}\label{delta-f-from-the-left}
	Let \(p \in Q\) and \(a \in \Sigma\). Then
		\[\defined\delta_f(p, a) \implies {\delta_4}_{f_4}(p, a) = \delta_f(p, a) \land {\lambda_4}_{f_4}(p, a) = \lambda_f(p, a).\]
	\begin{proof}
		We proceed by induction on \(\level_f(p)\).

		First, suppose \(\level_f(p) = 0\). Then \(\delta_f(p, a) = \delta(p, a)\) and \(\lambda_f(p, a) = \lambda(p, a)\). Let \(\langle p_1, p_2 \rangle \coloneqq p\) and \(\langle q_1, q_2 \rangle \coloneqq \delta(p, a)\).
		\begin{enumerate}
			\item Suppose \(\defined\delta_2(p_2, \omega)\), \(p_1, q_1 \in Q_\omega^l\) for some \(\omega \in \Omega\), \(\langle p_1, a, q_1 \rangle \in \delta_1\), and \(q_2 = p_2\). By \autoref{left-right-property}, \(\Proj_1(\lambda_1(p_1, a)) = \varepsilon\) and \(\lambda(p, a) = \Proj_2(\lambda_1(p_1, a))\). Then by the definition of \(\delta_4\) it follows that
			\begin{align*}
				{\delta_4}_{f_4}(p, a) &= \delta_4(p, a) = \langle q_1, p_2 \rangle = \delta(p, a) = \delta_f(p, a), \\
				{\lambda_4}_{f_4}(p, a) &= \lambda_4(p, a) = \Proj_2(\lambda_1(p_1, a)) = \lambda(p, a) = \lambda_f(p, a).
			\end{align*}
			
			\item Suppose \(\delta_2(p_2, \omega) = q_2\), \(\langle p_1, a, q_1' \rangle \in \Delta_\omega\) for some \(\omega \in \Omega\), and \(q_1 = E(q_1')\). Then by definition \(\langle p_1, a, q_1' \rangle \in \delta_1\), \(\Proj_1(\lambda_1(p_1, a)) = \omega\) and \(\lambda(p, a) = \Proj_2(\lambda_1(p_1, a)) \lambda_2(p_2, \omega)\). Also, \(\delta_3(p_1, a) = E(q_1')\). From the definition of \(\delta_4\) it follows that
			\begin{align*}
				{\delta_4}_{f_4}(p, a) &= \delta_4(p, a) = \langle E(q_1'), q_2 \rangle = \langle q_1, q_2 \rangle = \delta(p, a) = \delta_f(p, a), \\
				{\lambda_4}_{f_4}(p, a) &= \lambda_4(p, a) = \Proj_2(\lambda_1(p_1, a)) \lambda_2(p_2, \omega) = \lambda(p, a) = \lambda_f(p, a).
			\end{align*}
			
			\item Suppose \(\defined\delta_2(p_2, \omega)\), \(p_1, q_1 \in Q_\omega^r\) for some \(\omega \in \Omega\), \(\langle p_1, a, q_1 \rangle \in \delta_1\) and \(q_2 = p_2\). The reasoning is the same as in the first case.
		\end{enumerate}
		
		Now, suppose \(\level_f(p) > 0\). If \(\delta(p, a)\) is defined, the reasoning is the same as in the base case. Suppose \(\neg\defined\delta(p, a)\). Then \(\delta_f(p, a) = \delta_f(f(p), a) = {\delta_4}_{f_4}(f(p), a)\) and \(\lambda_f(p, a) = \lambda_f(f(p), a) = {\lambda_4}_{f_4}(f(p), a)\). Because \(f \subseteq f_4\) and using \autoref{delta-4-implies-delta}
		\begin{align*}
			{\delta_4}_{f_4}(f(p), a) &= {\delta_4}_{f_4}(f_4(p), a) = {\delta_4}_{f_4}(p, a), \\
			{\lambda_4}_{f_4}(f(p), a) &= {\lambda_4}_{f_4}(f_4(p), a) = {\lambda_4}_{f_4}(p, a).
		\end{align*}
	\end{proof}
\end{proposition}

\begin{definition}
	Let \(\mathcal{T}\) be a transducer or a failure transducer with alphabet \(\Sigma\), states \(Q\) and transition function \(\delta\). We define \(\Sig_\mathcal{T} \colon Q \to 2^\Sigma\) for \(q \in Q\) as
	\[\Sig_\mathcal{T}(q) \coloneqq \{a \in \Sigma \mid \defined\delta(q, a)\}.\]
\end{definition}

\begin{proposition}\label{f-4-unnecessary-failure-aux}
	Let \(\langle p_2, \omega \rangle \in \Dom(\delta_2)\) and \(p_1 \in Q_\omega^r\). Then
	\[\defined f_4(\langle p_1, p_2 \rangle) \implies \Sig_{\mathcal{V}^* \circ \mathcal{F}}(\langle p_1, p_2 \rangle) = \Sig_{\mathcal{V}^* \circ \mathcal{F}}(f_4(\langle p_1, p_2 \rangle)).\]
	\begin{proof}
		By definition \(f_4(\langle p_1, p_2 \rangle) = \langle p_1, f_2(p_2) \rangle\). Since \(\mathcal{V}^* \circ \mathcal{F}\) is monotonic, \(\Sig_{\mathcal{V}^* \circ \mathcal{F}}(\langle p_1, p_2 \rangle) \subseteq \Sig_{\mathcal{V}^* \circ \mathcal{F}}(\langle p_1, f_2(p_2) \rangle)\). Suppose \(a \in \Sig_{\mathcal{V}^* \circ \mathcal{F}}(\langle p_1, f_2(p_2) \rangle)\). From \autoref{generic-composition} it follows that \(\defined\delta_1(p_1, a)\). Also, \(\Proj_1(\lambda_1(p_1, a)) = \varepsilon\) because \(p_1 \in Q_\omega^r\). Therefore, \(\defined\delta_4(\langle p_1, p_2 \rangle, a)\), i.e. \(a \in \Sig_{\mathcal{V}^* \circ \mathcal{F}}(\langle p_1, p_2 \rangle)\).
	\end{proof}
\end{proposition}	

\begin{proposition}\label{f-4-unnecessary-failure}
	Let \(\langle p_2, \omega \rangle \in \Dom(\delta_2)\),  \(p_1 \in Q_\omega^r\) and \(\langle p_1, p_2 \rangle \in Q\). Then
	\[(\forall a \in \Sigma)(\neg\defined\delta_4(\langle p_1, p_2 \rangle, a) \land \defined f_4(\langle p_1, p_2 \rangle) \implies \neg\defined{\delta_4}_{f_4}(\langle p_1, p_2 \rangle, a)).\]
	\begin{proof}
		Follows directly from \autoref{f-4-unnecessary-failure-aux} with induction on the number of failure transitions.
	\end{proof}
\end{proposition}	

\begin{proposition}\label{delta-f-from-the-right}
	Let \(p \in Q\), \(a \in \Sigma\) be such that \(\defined{\delta_4}_{f_4}(p, a)\) and there exist \(\alpha, \alpha' \in \Sigma^*\) such that \({\delta_4}_{f_4}^*(s_4, \alpha) = p\), \({\delta_4}_{f_4}^*(p, a\alpha') \in F_4\). Then \(\delta_f(p, a) = {\delta_4}_{f_4}(p, a)\).
	\begin{proof}
		We proceed by induction on \(\level_{f_4}(p)\).

		First, suppose \(\level_{f_4}(p) = 0\). Then \({\delta_4}_{f_4}(p, a) = \delta_4(p, a)\). Let \(\langle p_1, p_2 \rangle \coloneqq p\) and \(\langle q_1, q_2 \rangle \coloneqq \delta_4(p, a)\). Since \({\delta_4}_{f_4}^*(\langle q_1, q_2 \rangle, \alpha') \in F_4\), \(F_3 = \{s_3\}\) and \(\rho_3(s_3) = \langle \varepsilon, 1 \rangle\), \(F_4 = \{s_1\} \times F_2\) and there exists \(\beta'\) -- prefix of \(\alpha'\), such that \({\delta_4}_{f_4}^*(\langle q_1, q_2 \rangle, \beta') = \langle s_1, r_1 \rangle\). Let \(\beta'\) be the shortest prefix of \(\alpha'\) such that \(\Proj_1({\delta_4}_{f_4}^*(\langle q_1, q_2 \rangle, \beta')) = s_1\). Let \(\gamma\) be the longest prefix of \(\alpha\) (\(\alpha = \gamma\beta\)), such that \(\Proj_1({\delta_4}_{f_4}^*(\langle s_1, s_2 \rangle, \gamma)) = s_1\). This means that \(\langle s_1, \beta a\beta', s_1 \rangle \in \delta_3\) without going through an intermediate state \(s_1\). Therefore, from \autoref{kleene-star-auxiliary2} \((\exists f_1 \in F_1)(\delta_1^*(s_1, \beta a\beta') = f_1)\). Let \(\omega \coloneqq \lambda_1^*(s_1, \beta a\beta')\), \(m \coloneqq |\beta a\beta'|\), \(t_1 \coloneqq s_1\), for \(1 \le i \le m\), \(t_{i+1} \coloneqq \delta_1(t_i, b_i)\), where \(b_i\) is the \(i\)-th symbol of \(\beta a\beta'\), and \(t_{m+1} = f_1\). Then there exists \(1 \le i \le m\) such that \((\forall 1 \le j \le i)(t_j \in Q_\omega^l)\) and \((\forall i < j \le m+1)(t_j \in Q_\omega^r)\). Let \(t_j = p_1\).
		\begin{enumerate}
			\item Suppose \(j < i\). Then \(p_1, q_1 \in Q_\omega^l\) and \(p_1, q_1 \notin F_1\), i.e. \(\langle p_1, a, q_1 \rangle \in \delta_1\) and \(\Proj_1(\lambda_1(p_1, a)) = \varepsilon\). By \autoref{generic-composition}, \(\langle p_1, a, q_1 \rangle \in \delta_3\), \(\Proj_1(\lambda_3(p_1, a)) = \Proj_1(\lambda_1(p_1, a)) = \varepsilon\) and \(q_2 = p_2\). By \autoref{optimized-composition}, \(\delta(\langle p_1, p_2 \rangle, a) = \langle q_1, p_2 \rangle = \langle q_1, q_2 \rangle\).
			\item Suppose \(j = i\). Then \(p_1 \in Q_\omega^l\), \(t_{j+1} \in Q_\omega^r\) and \(\langle p_1, a, t_{j+1} \rangle \in \Delta_\omega\). Since \(p_1 \in Q_\omega^l\), we have that \(p_1 \not\in F_1\) and therefore \(\langle p_1, a, E(t_{j+1}) \rangle = \langle p_1, a, q_1 \rangle\). By \autoref{generic-composition}, \(\langle p_1, a, q_1 \rangle \in \delta_3\), \(\omega = \Proj_1(\lambda_3(p_1, a)) = \Proj_1(\lambda_1(p_1, a))\) and \(\langle p_2, \omega, q_2 \rangle \in \delta_2\). By \autoref{optimized-composition}, \(\delta(\langle p_1, p_2 \rangle, a) = \langle E(t_{j+1}), q_2 \rangle = \langle q_1, q_2 \rangle\).
			\item Suppose \(j > i\). Then \(p_1, t_{j+1} \in Q_\omega^r\) and \(\Proj_1(\lambda_1(p_1, a)) = \varepsilon\). By \autoref{generic-composition}, \(\langle p_1, a, q_1 \rangle \in \delta_3\), \(\Proj_1(\lambda_3(p_1, a)) = \Proj_1(\lambda_1(p_1, a)) = \varepsilon\), \(q_1 = E(t_{j+1})\) and \(q_2 = p_2\). \(p_1 \notin F_1\), therefore \(E(p_1) = p_1\) and \(\neg(\exists \omega \in \Omega)(E(p_1) \in Q_\omega^l)\). Since \(\langle p_1, p_2 \rangle \in Q\), exists \(\langle l_2, \omega, p_2 \rangle \in \delta_2\). Thus, by \autoref{optimized-composition}, \(\delta(\langle p_1, p_2 \rangle, a) = \langle E(t_{j+1}), p_2 \rangle = \langle q_1, q_2 \rangle\).
		\end{enumerate}
		
		Now, suppose \(\level_{f_4}(p) > 0\). If \(\delta_4(p, a)\) is defined, the reasoning is the same as in the base case. Suppose \(\neg\defined\delta_4(p, a)\). Then \({\delta_4}_{f_4}(p, a) = {\delta_4}_{f_4}(f_4(p), a)\). If \(p \notin \Dom(f)\), then \(p_1 \in Q_\omega^r\) and by \autoref{f-4-unnecessary-failure} \(\neg\defined\delta_{f_4}(p, a)\), which contradicts our assumption. Therefore, \(p \in \Dom(f)\) and since \(f \subseteq f_4\), \({\delta_4}_{f_4}(f_4(p), a) = {\delta_4}_{f_4}(f(p), a) = \delta_f(f(p), a) = \delta_f(p, a)\).
	\end{proof}
\end{proposition}

\begin{proposition}\label{optimized-composition-correctness-aux}
	\((\forall \langle q_1, q_2 \rangle \in Q) (\mathcal{O}_{\mathcal{W}}^{\langle q_1, q_2 \rangle} = \mathcal{O}_{\mathcal{V}^*\circ \mathcal{F}}^{\langle q_1, q_2 \rangle})\)
	
	\begin{proof}
		Let \(\alpha \in \Dom(\mathcal{O}_\mathcal{W}^{\langle q_1, q_2 \rangle})\), \(\alpha = a_1 a_2 \ldots a_n\) and \((\forall 1 \le i \le n)(a_i \in \Sigma)\). Then there exist \(p_1, p_2, \ldots, p_{n+1} \in Q\), such that \(\delta_f(p_i, a_i) = p_{i+1}\) for \(1 \le i \le n\), \(p_1 = \langle q_1, q_2 \rangle\) and \(p_{n+1} = \langle s_1, t_2 \rangle\), where \(t_2 \in F_2\). \autoref{delta-f-from-the-left} gives us that \({\delta_4}_{f_4}(p_i, a_i) = p_{i+1}\) and \({\lambda_4}_{f_4}(p_i, a_i) = \lambda_f(p_i, a_i)\) for \(1 \le i \le n\). Since \(s_3 = s_1\), we get \(p_{n+1} = \langle s_3, t_2 \rangle \in F_4\). Therefore, \(\alpha \in \Dom(\mathcal{O}_{\mathcal{V}^* \circ \mathcal{F}}^{\langle q_1, q_2 \rangle} )\). Using \(\rho_1(s_3) = \langle \varepsilon, 1 \rangle\) we obtain
		\begin{align*}
			\mathcal{O}_{\mathcal{V}^* \circ \mathcal{F}}^{\langle q_1, q_2 \rangle}(\alpha) ={}& {\lambda_4}_{f_4}^*(\langle q_1, q_2 \rangle, \alpha) \rho_4(\langle s_3, t_2 \rangle) \\
			={}& {\lambda_4}_{f_4}^*(\langle q_1, q_2 \rangle, \alpha) \\
			&[\Proj_2(\rho_1(s_3)) {\lambda_2}_{f_2}^*(t_2, \Proj_1(\rho_1(s_3))) \rho_2({\delta_2}_{f_2}^*(t_2, \Proj_1(\rho_1(s_3)))] \\
			={}& {\lambda_4}_{f_4}^*(\langle q_1, q_2 \rangle, \alpha) [1 {\lambda_2}_{f_2}^*(t_2, \varepsilon) \rho_2({\delta_2}_{f_2}^*(t_2, \varepsilon))] \\
			={}& {\lambda_4}_{f_4}^*(\langle q_1, q_2 \rangle, \alpha) \rho_2(t_2) \\
			={}& \lambda_f^*(\langle q_1, q_2 \rangle, \alpha) \rho(\delta_f^*(\langle q_1, q_2 \rangle, \alpha)) \\
			={}&\mathcal{O}_\mathcal{W}^{\langle q_1, q_2 \rangle}(\alpha).
		\end{align*}
		
		Let \(\alpha \in \Dom(\mathcal{O}_{\mathcal{V}^* \circ \mathcal{F}}^{\langle q_1, q_2 \rangle})\), \(\alpha = a_1 a_2 \ldots a_n\) and \((\forall 1 \le i \le n)(a_i \in \Sigma)\). In other words, there exist \(p_1, p_2, \ldots, p_{n+1} \in Q\), such that \({\delta_4}_{f_4}(p_i, a_i) = p_{i+1}\) for \(1 \le i \le n\), \(p_1 = \langle q_1,q_2 \rangle\) and \(p_{n+1} = \langle s_3, t_2 \rangle\), where \(t_2 \in F_2\). \autoref{delta-f-from-the-right} gives us that \(\delta_f(p_i, a_i) = p_{i+1}\) and \(\lambda_f(p_i, a_i) = {\lambda_4}_{f_4}(p_i, a_i)\) for \(1 \le i \le n\). Since \(s_1 = s_3\), we get that \(p_{n+1} = \langle s_1, t_2 \rangle\) is final in \(\mathcal{W}\). Therefore, \(\alpha \in \Dom(\mathcal{O}_\mathcal{W}^{\langle q_1, q_2 \rangle} )\).
	\end{proof}
\end{proposition}

\begin{proposition}\label{W-equivalent-to-generic-composition}
	\(\mathcal{O}_{\mathcal{W}} = \mathcal{O}_{\mathcal{V}^*\circ \mathcal{F}}\).
\begin{proof}
	By applying \autoref{optimized-composition-correctness-aux} for the initial state \(\langle s_1, s_2 \rangle\) and using the fact that \(s_1 = s_3\), we obtain
	\[\Dom(\mathcal{O}_{\mathcal{W}}) = \Dom(\mathcal{O}_{\mathcal{W}}^{\langle s_1, s_2 \rangle}) = \Dom(\mathcal{O}_{\mathcal{V}^* \circ \mathcal{F}}^{\langle s_1, s_2 \rangle}) = \Dom(\mathcal{O}_{\mathcal{V}^* \circ \mathcal{F}}).\]
	For \(\alpha \in \Dom(\mathcal{O}_{\mathcal{W}})\), using that \(\iota_1 = \langle \varepsilon, 1 \rangle\), we obtain

	\begin{align*}
		\mathcal{O}_{\mathcal{V}^* \circ \mathcal{F}}(\alpha) ={}& \iota_4 \mathcal{O}_{\mathcal{V}^* \circ \mathcal{F}}^{\langle s_1, s_2 \rangle}(\alpha) \\
		={}& [\Proj_2(\iota_1) \iota_2 {\lambda_2}_{f_2}^*(s_2, \Proj_1(\iota_1))] \mathcal{O}_{\mathcal{V}^* \circ \mathcal{F}}^{\langle s_1, s_2 \rangle}(\alpha)  \\
		={}& [1 \iota_2 {\lambda_2}_{f_2}^*(s_2, \varepsilon)] \mathcal{O}_{\mathcal{V}^* \circ \mathcal{F}}^{\langle s_1, s_2 \rangle}(\alpha)  \\
		={}& \iota_2 \mathcal{O}_\mathcal{W}^{\langle s_1, s_2 \rangle}(\alpha) \\
		={}& \mathcal{O}_\mathcal{W}(\alpha).
	\end{align*}
\end{proof}
\end{proposition}

\begin{proposition}\label{W-second-coordinate-delta-f}
	For every \(\langle p, \omega \rangle \in \delta_2\) and \(\alpha \in \Dom(\mathcal{O}_{\mathcal{V}^*}(\bullet \mid \omega))\),
	\[\delta_f^*(\langle s_1, p \rangle, \alpha) = \langle s_1, {\delta_2}_{f_2}(p, \omega)\rangle.\]

	\begin{proof}
		\(s_1 \in Q_\omega^l\) and \(s_3 = s_1\). Then using \autoref{delta-f-from-the-left} we conclude that \(\delta_f^*(\langle s_1, p \rangle, \alpha) = {\delta_4}_{f_4}^*(\langle s_1, p \rangle, \alpha)\). From \autoref{generic-composition-delta-f-star} it follows that
		\[{\delta_4}_{f_4}^*(\langle s_1, p \rangle, \alpha) = \langle \delta_3^*(s_1, \alpha), {\delta_2}_{f_2}^*(p, \omega) \rangle = \langle s_1, {\delta_2}_{f_2}(p, \omega) \rangle.\]
	\end{proof}
\end{proposition}

\begin{proposition}\label{W-second-coordinate-delta-f-star}
	For every \( p \in Q_2\), \(\omega \in \Omega^*\) and \(\alpha \in \Dom(\mathcal{O}_{\mathcal{V}^*}(\bullet \mid \omega))\),
	\[\delta_f^*(\langle s_1, p \rangle, \alpha) = \langle s_1, {\delta_2}_{f_2}^*(p, \omega)\rangle.\]
	\begin{proof}
		Follows from \autoref{W-second-coordinate-delta-f} by induction on the length of \(\omega\).
	\end{proof}
\end{proposition}

\begin{proposition}
	Every state in \(\mathcal{W}\) is co-accessible.
	\begin{proof}
		Let \(\langle p_2, \omega \rangle \in \delta_2\) and \(p_1 \in Q_\omega^r\) such that \(\langle p_1, p_2 \rangle \in Q\). Since every state in \(\mathcal{V}\) is co-accessible, there exist \(\alpha \in \Sigma^*\) and \(q_1 \in F_1\) such that \(\delta_1^*(p_1, \alpha) = q_1\). \(\lambda_1^*(p_1, \alpha) = \varepsilon\) because \(p_1 \in Q_\omega^r\). Therefore, \(\delta^*(\langle p_1, p_2 \rangle, \alpha) = \langle s_1, p_2 \rangle\). Every state in \(\mathcal{F}\) is also co-accessible. Therefore, \((\exists \beta \in \Omega^*)(\exists q_2 \in F_2)({\delta_2}_{f_2}^*(p_2, \beta) = q_2)\). Let \(\gamma \in \Dom(\mathcal{O}_{\mathcal{V}^*}(\bullet \mid \beta))\). By \autoref{W-second-coordinate-delta-f-star}, \(\delta_f^*(\langle s_1, p_2 \rangle, \gamma) = \langle s_1, q_2 \rangle \in F\). In other words, \(\delta_f^*(\langle p_1, p_2 \rangle, \alpha\gamma) \in F\).
		
		Let \(\langle p_2, \omega \rangle \in \delta_2\) and \(p_1 \in Q_\omega^l\) such that \(\langle p_1, p_2 \rangle \in Q\). Then there exists \(\langle l_1, a, r_1 \rangle \in \Delta_\omega\), \((\exists \alpha \in \Sigma^*)(\delta_1^*(p_1, \alpha) = l_1)\) and \(\Proj_1(\lambda_1^*(p_1, \alpha)) = \varepsilon\). This implies that \(\delta^*(\langle p_1, p_2 \rangle, \alpha) = \langle l_1, p_2 \rangle\). By the definition of \(\delta\), we have that \(\delta(\langle l_1, p_2 \rangle, a) = \langle E(r_1), \delta_2(p_2, \omega) \rangle\), i.e. \(\delta^*(\langle p_1, p_2 \rangle, \alpha a) = \langle E(r_1), \delta_2(p_2, \omega) \rangle\). If \(E(r_1) = s_1\), then we apply \autoref{W-second-coordinate-delta-f-star}, otherwise \(r_1 \in Q_\omega^r\) and we reason as in the first case.
	\end{proof}
\end{proposition}

\begin{proposition}\label{W-monotonic-and-probabilistic}
	\(\mathcal{W}\) is monotonic and probabilistic.
	\begin{proof}
		Let \(p \in Q\) and \(\defined f(p)\). Then \(\defined f_4(p)\) and \(f_4(p) = f(p)\). If \(p \in \{s_1\} \times F_2\), then \(p \in F_4\) and because \(\mathcal{V}^* \circ \mathcal{F}\) is monotonic, \(f_4(p) = f(p) \in F_4 = \{s_1\} \times F_2\). Let \(\langle p, a \rangle \in \Dom(\delta)\). Then by the base case of \autoref{delta-f-from-the-left} we obtain that \(\langle p, a \rangle \in \Dom(\delta_4)\). \(\mathcal{V}^* \circ \mathcal{F}\) is monotonic. Thus, \(\langle f_4(p), a \rangle = \langle f(p), a \rangle \in \Dom(\delta_4)\) and by \autoref{delta-4-implies-delta}, \(\langle f(p), a \rangle \in \Dom(\delta)\).
	
		\autoref{kleene-star-correctness} implies that \(\mathcal{V}^* \circ \mathcal{F}\) is probabilistic. \(\mathcal{O}_\mathcal{W} = \mathcal{O}_{\mathcal{V}^* \circ \mathcal{F}}\) by \autoref{W-equivalent-to-generic-composition}. Therefore, \(\mathcal{W}\) is also probabilistic.
	\end{proof}
\end{proposition}

\section{Weight Pushing}

Let \(\mathcal{V}\) and \(\mathcal{F}\) be as described in \autoref{special-composition}. Also, let \(\mathcal{F}\) be stochastic and \(\mathcal{W} \coloneqq \langle \Sigma, \mathcal{R}, Q, s, F,\allowbreak \delta, \lambda, \iota, \rho, f, \varphi \rangle\) be the failure transducer from \autoref{optimized-composition}, equivalent to the composition of \(\mathcal{V}^*\) and \(\mathcal{F}\). From \autoref{W-monotonic-and-probabilistic} it follows that \(\mathcal{W}\) is monotonic and probabilistic but it is not necessarily stochastic.

For a probabilistic transducer ${\cal T}$, a semiring \(\mathcal{S} \coloneqq \langle \mathbb{R}_+, \oplus, \times, 0, 1 \rangle\), and a state \(q\) with $S_{\cal T}(q)$ we denote the sum
$\bigoplus_{\alpha \in \Dom(\mathcal{O}_\mathcal{T}^q)} \mathcal{O}_\mathcal{T}^q(\alpha).$
Since \(\mathcal{W}\) is probabilistic, the sums \(S_\mathcal{W}(q)\) exist for every \(q \in Q\) and the following construction can be used to obtain the canonical form of \(\mathcal{W}\).

\begin{definition}\label{weight-pushing-construction}
	Let \(\mathcal{W}_\mathcal{C} \coloneqq \langle \Sigma, \mathcal{R}, Q, s, F, \delta, \lambda_\mathcal{C}, \iota_\mathcal{C}, \rho_\mathcal{C}, f, \varphi_\mathcal{C} \rangle\), where
	\begin{itemize}
		\item \(\lambda_\mathcal{C} \coloneqq \{\langle p, a, \frac{w S_\mathcal{W}(\delta(p, a))}{S_\mathcal{W}(p)} \rangle \mid \langle p, a, w \rangle \in \lambda\}\);
		\item \(\iota_\mathcal{C} \coloneqq \iota S_\mathcal{W}(s)\);
		\item \(\rho_\mathcal{C} \coloneqq \{\langle p, \frac{w}{S_\mathcal{W}(p)} \rangle \mid \langle p, w\rangle \in \rho\}\);
		\item \(\varphi_\mathcal{W} \coloneqq \{\langle p, \frac{w S_\mathcal{W}(f(p))}{S_\mathcal{W}(p)} \rangle \mid \langle p, w \rangle \in \varphi\}\).
	\end{itemize}
\end{definition}

\begin{remark}
	\(\mathcal{W}\) and \(\mathcal{W}_\mathcal{C}\) have the same states, final states, transitions and failure transitions, therefore \(\mathcal{W}_\mathcal{C}\) has no failure cycles and \(\Dom(\mathcal{O}_\mathcal{W}^p) = \Dom(\mathcal{O}_{\mathcal{W}_\mathcal{C}}^p)\) for every \(p \in Q\).
\end{remark}

\begin{proposition}
	Let \(p \in Q\) and \(a \in \Sigma^*\) be such that \(\defined\delta_f(p, a)\). Then
	\[{\lambda_\mathcal{C}}_f(p, a) = \frac{\lambda_f(p, a) S_\mathcal{W}(\delta_f(p, a))}{S_\mathcal{W}(p)}.\]
	\begin{proof}
		We proceed by induction on \(\level_{f_\mathcal{C}}(p)\).
		
		First, suppose \(\level_{f_\mathcal{C}}(p) = 0\). Then \(\delta_f(p, a) = \delta(p, a)\) and
		\[{\lambda_\mathcal{C}}_f(p, a) = \lambda_\mathcal{C}(p, a) = \frac{\lambda(p, a) S_\mathcal{W}(\delta(p, a))}{S_\mathcal{W}(p)} = \frac{\lambda_f(p, a) S_\mathcal{W}(\delta_f(p, a))}{S_\mathcal{W}(p)}.\]
		
		Now, suppose \(\level_{f_\mathcal{C}}(p) > 0\). If \(\defined\delta(p, a)\), then the reasoning is the same as in the base case. Suppose \(\neg\defined\delta(p, a)\). Then
		\begin{align*}
			{\lambda_\mathcal{C}}_f(p, a) &= \varphi_\mathcal{C}(p){\lambda_\mathcal{C}}_f(f(p), a) \\
			&= \frac{\varphi(p) S_\mathcal{W}(f(p))}{S_\mathcal{W}(p)} \frac{\lambda_f(f(p), a) S_\mathcal{W}(\delta_f(f(p), a))}{S_\mathcal{W}(f(p))} \\
			&= \frac{\lambda_f(p, a) S_\mathcal{W}(\delta_f(p, a))}{S_\mathcal{W}(p)}.
		\end{align*}
	\end{proof}
\end{proposition}

\begin{proposition}
	Let \(p \in Q\) and \(\alpha \in \Sigma^*\) be such that \(\defined\delta_f^*(p, \alpha)\). Then
	\[{\lambda_\mathcal{C}}_f^*(p, \alpha) = \frac{\lambda_f^*(p, \alpha) S_\mathcal{W}(\delta_f^*(p, \alpha))}{S_\mathcal{W}(p)}.\]
	\begin{proof}
		We proceed by induction on \(|\alpha|\).
		
		First, suppose \(\alpha = \varepsilon\). Then
		\[{\lambda_\mathcal{C}}_f^*(p, \varepsilon) = 1 = \frac{\lambda_f^*(p, \varepsilon) S_\mathcal{W}(\delta_f^*(p, \varepsilon))}{S_\mathcal{W}(p)}.\]
		
		Now, suppose \(\alpha = \alpha' a\). Then
		\begin{align*}
			{\lambda_\mathcal{C}}_f^*(p, \alpha' a) &= {\lambda_\mathcal{C}}_f^*(p, \alpha') {\lambda_\mathcal{C}}_f(\delta_f^*(p, \alpha'), a) \\
			&= \frac{\lambda_f^*(p, \alpha') S_\mathcal{W}(\delta_f^*(p, \alpha'))}{S_\mathcal{W}(p)} \frac{\lambda_f(\delta_f^*(p, \alpha'), a) S_\mathcal{W}(\delta_f(\delta_f^*(p, \alpha'), a))}{S_\mathcal{W}(\delta_f^*(p, \alpha'))} \\
			&= \frac{\lambda_f^*(p, \alpha) S_\mathcal{W}(\delta_f^*(p, \alpha))}{S_\mathcal{W}(p)}.
		\end{align*}
	\end{proof}
\end{proposition}

\begin{proposition}
	Let \(p \in Q\) and \(\alpha \in \Dom(\mathcal{O}_\mathcal{W}^p)\). Then
	\[\mathcal{O}_{\mathcal{W}_\mathcal{C}}^p(\alpha) = \frac{\mathcal{O}_\mathcal{W}^p(\alpha)}{S_\mathcal{W}(p)}.\]
	\begin{proof}
		\begin{align*}
			\mathcal{O}_{\mathcal{W}_\mathcal{C}}^p(\alpha) &= {\lambda_\mathcal{C}}_f^*(p, \alpha) \rho_\mathcal{C}(\delta_f^*(p, \alpha)) \\
			&= \frac{\lambda_f^*(p, \alpha) S_\mathcal{W}(\delta_f^*(p, \alpha))}{S_\mathcal{W}(p)} \frac{\rho(\delta_f^*(p, \alpha))}{S_\mathcal{W}(\delta_f^*(p, \alpha))} \\
			&= \frac{\lambda_f^*(p, \alpha) \rho(\delta_f^*(p, \alpha))}{S_\mathcal{W}(p)} \\
			&= \frac{\mathcal{O}_\mathcal{W}^p(\alpha)}{S_\mathcal{W}(p)}.
		\end{align*}
	\end{proof}
\end{proposition}

\begin{proposition}
	\(\mathcal{O}_{\mathcal{W}_\mathcal{C}} = \mathcal{O}_\mathcal{W}\).
	\begin{proof}
		Let \(\alpha \in \Dom(\mathcal{O}_\mathcal{W})\). Then
		\begin{align*}
			\mathcal{O}_{\mathcal{W}_\mathcal{C}}(\alpha) = \iota_\mathcal{C} \mathcal{O}_{\mathcal{W}_\mathcal{C}}^s(\alpha) = \iota S_\mathcal{W}(s) \frac{\mathcal{O}_\mathcal{W}^s(\alpha)}{S_\mathcal{W}(s)} = \iota \mathcal{O}_\mathcal{W}^s(\alpha) = \mathcal{O}_\mathcal{W}(\alpha).
		\end{align*}
	\end{proof}
\end{proposition}

\begin{proposition}
	\(\mathcal{W}_\mathcal{C}\) is canonical with respect to \(\mathcal{S}\).
	\begin{proof}
		Let \(p \in Q\). Then
		\begin{align*}
			\bigoplus_{\alpha \in \Dom(\mathcal{O}_{\mathcal{W}_\mathcal{C}}^p)} \mathcal{O}_{\mathcal{W}_\mathcal{C}}^p(\alpha) &= \bigoplus_{\alpha \in \Dom(\mathcal{O}_\mathcal{W}^p)} \frac{\mathcal{O}_\mathcal{W}^p(\alpha)}{S_\mathcal{W}(p)} \\
			&= \frac{1}{S_\mathcal{W}(p)} \bigoplus_{\alpha \in \Dom(\mathcal{O}_\mathcal{W}^p)} \mathcal{O}_\mathcal{W}^p(\alpha) \\
			&= \frac{1}{S_\mathcal{W}(p)} S_\mathcal{W}(p) \\
			&= 1.
		\end{align*}
	\end{proof}
\end{proposition}

\begin{remark}
	Since \(\mathcal{F}\) is stochastic, it is also canonical with respect to \(\mathcal{R}^+\) (see \autoref{stochastic-property}), i.e. for every state \(q\) of \(\mathcal{F}\), \(S_\mathcal{F}(q) = 1\).
\end{remark}

\subsection{Weight Pushing in \(\mathcal{R}^+\)}

\begin{proposition}\label{prop-8}
	\((\forall p \in Q_2)\left(S_\mathcal{W}(\langle s_1, p \rangle) = 1\right)\).
	\begin{proof}
		Using \autoref{generic-composition-correctness-aux} and the fact that \(\iota_1 = \langle \varepsilon, 1 \rangle\)
		\begin{align*}
			S_\mathcal{W}(\langle s_1, p \rangle) &= \sum_{\alpha \in \Dom(\mathcal{O}_\mathcal{W}^{\langle s_1, p \rangle})} \mathcal{O}_\mathcal{W}^{\langle s_1, p \rangle}(\alpha) \\
			&= \sum_{\alpha \in \Dom(\mathcal{O}_{\mathcal{V}^* \circ \mathcal{F}}^{\langle s_1, p \rangle})} \mathcal{O}_{\mathcal{V}^* \circ \mathcal{F}}^{\langle s_1, p \rangle}(\alpha) \\
			&= \sum_{\beta \in \Dom(\mathcal{O}_\mathcal{F}^p)} \mathcal{O}_\mathcal{F}^p(\beta) \sum_{\alpha \in \Dom(\mathcal{O}_{\mathcal{V}^*}^{s_1}(\bullet \mid \beta))} \mathcal{O}_{\mathcal{V}^*}^{s_1}(\alpha \mid \beta) \\
			&= \sum_{\beta \in \Dom(\mathcal{O}_\mathcal{F}^p)} \mathcal{O}_\mathcal{F}^p(\beta) \sum_{\alpha \in \Dom(\mathcal{O}_{\mathcal{V}^*}(\bullet \mid \beta))} \mathcal{O}_{\mathcal{V}^*}(\alpha \mid \beta) \\
			&= \sum_{\beta \in \Dom(\mathcal{O}_\mathcal{F}^p)} \mathcal{O}_\mathcal{F}^p(\beta) \\
			&= 1.
		\end{align*}
	\end{proof}
\end{proposition}

\begin{definition}
	Let \(\widetilde{\mathcal{W}} \coloneqq \langle \Sigma, \mathcal{R}, Q, s, Q_{s_1}, \widetilde{\delta},\allowbreak \widetilde{\lambda}, \iota, Q_{s_1} \times \{1\}, \widetilde{f}, \widetilde{\varphi} \rangle\), where
	\begin{itemize}
		\item \(Q_{s_1} \coloneqq \{\langle p, q \rangle \in Q \mid p = s_1\}\) and \(\overline{Q}_{s_1} \coloneqq Q \setminus Q_{s_1}\);
		\item \(\widetilde{\delta} \coloneqq \delta\restriction_{\overline{Q}_{s_1} \times \Sigma}\), \(\widetilde{\lambda} \coloneqq \lambda\restriction_{\overline{Q}_{s_1} \times \Sigma}\), \(\widetilde{f} \coloneqq f\restriction_{\overline{Q}_{s_1} \times \Sigma}\), and \(\widetilde{\varphi} \coloneqq \varphi\restriction_{\overline{Q}_{s_1} \times \Sigma}\).
	\end{itemize}
\end{definition}

\begin{proposition}
	\((\forall q \in Q)(S_{\widetilde{\mathcal{W}}}(q) = S_{\mathcal{W}}(q))\).
	\begin{proof}
		Let \(q \in Q_{s_1}\). Then \(\Dom(\mathcal{O}_{\widetilde{\mathcal{W}}}^q) = \{\varepsilon\}\) and \(S_{\widetilde{\mathcal{W}}}(q) = \widetilde{\rho}(q) = 1 = S_\mathcal{W}(q)\).
		
		Let \(q \in \overline{Q}_{s_1}\). \(\Dom(\mathcal{O}_{\widetilde{\mathcal{W}}}^q)\) is the set of all prefixes \(\alpha'\) of \(\Dom(\mathcal{O}_\mathcal{W}^q)\) such that \(\delta^*(q, \alpha') \in Q_{s_1}\) without going through an intermediate state from \(Q_{s_1}\). By definition for every \(\alpha' \in \Dom(\mathcal{O}_{\widetilde{\mathcal{W}}}^q)\) it holds that \(\lambda^*(q, \alpha') = \widetilde{\lambda}^*(q, \alpha')\). Thus,
		\begin{align*}
			S_\mathcal{W}(q) &= \sum_{\alpha \in \Dom(\mathcal{O}_\mathcal{W}^q)} \mathcal{O}_\mathcal{W}^q(\alpha) \\
			&\stackrel{\mathclap{\normalfont\mbox{\tiny{\autoref{prop-8}}}}}{=} \sum_{\alpha' \in \Dom(\mathcal{O}_{\widetilde{\mathcal{W}}}^q)} \lambda^*(q, \alpha') S_\mathcal{W}(\delta^*(q, \alpha')) \\
			&= \sum_{\alpha' \in \Dom(\mathcal{O}_{\widetilde{\mathcal{W}}}^q)} \widetilde{\lambda}^*(q, \alpha') \\
			&= \sum_{\alpha' \in \Dom(\mathcal{O}_{\widetilde{\mathcal{W}}}^q)} \mathcal{O}_{\widetilde{\mathcal{W}}}^q(\alpha') \\
			&= S_{\widetilde{\mathcal{W}}}(q).
		\end{align*}
	\end{proof}
\end{proposition}

\begin{definition}\label{cloned-states-graph}
	Let \(\mathcal{G} \coloneqq (V, E)\), where
	\begin{align*}
		V \coloneqq{}& Q \cup \{\langle q, \widetilde{f}(q) \rangle \mid q \in \Dom(\widetilde{f})\}, \\
		E \coloneqq{}& \{\langle p, \langle a, \widetilde{\lambda}(p, a) \rangle, q \rangle \mid \langle p, a, q \rangle \in \widetilde{\delta}\} \cup \\
		&\{\langle p, \langle \varepsilon, \widetilde{\varphi}(p) \rangle, \langle p, q \rangle \rangle \mid \langle p, q \rangle \in \widetilde{f}\} \cup \\
		&\{\langle \langle p, q \rangle, \langle a, \widetilde{\lambda}(q, a) \rangle, r \rangle \mid \langle p, q \rangle \in \widetilde{f}, \langle q, a, r \rangle \in \widetilde{\delta}, \neg\defined\widetilde{\delta}(p, a)\} \cup \\
	&\{\langle \langle p, q \rangle, \langle \varepsilon, \widetilde{\varphi}(q) \rangle, \langle q, r \rangle \rangle \mid \langle p, q \rangle \in \widetilde{f}, \langle q, r \rangle \in \widetilde{f}\}.
	\end{align*}
\end{definition}

For each path \(\pi\coloneqq \langle p_0, \langle l_1, \omega_1 \rangle, p_1, \langle l_2, \omega_2 \rangle, \ldots, p_n\rangle\) in \(\mathcal{G}\) we define

\[w(\pi) \coloneqq \prod_{i=1}^n  \omega_i,\]
\[l(\pi) \coloneqq l_1 l_2 \ldots l_n.\]

\begin{proposition}\label{delta-f-to-path}
	Let \(p \in Q\), \(a \in \Sigma\) and \(q \coloneqq \widetilde{\delta}_{\widetilde{f}}(p, a)\). Then there exists a unique path \(\pi\coloneqq \langle p_0, \langle l_1, \omega_1\rangle, p_1,  \langle l_2, \omega_2\rangle, \ldots, p_n \rangle\), where \(n > 0\), in \(\mathcal{G}\) such that \((\forall 1 \le i < n)(p_i \notin Q)\), \(p_0 = p\), \(p_n = q\), \(l(\pi) = a\) and \(w(\pi) = \widetilde{\lambda}_{\widetilde{f}}(p, a)\).
	\begin{proof}
		We proceed by induction on \(\level_f(p)\).
		
		First, suppose \(\level_{\widetilde{f}}(p) = 0\). Then \(q = \widetilde{\delta}_{\widetilde{f}}(p, a) = \widetilde{\delta}(p, a)\). By \autoref{cloned-states-graph}, \(\langle p, \langle a, \widetilde{\lambda}(p, a) \rangle, q \rangle \in E\).
		
		Now, suppose \(\level_{\widetilde{f}}(p) > 0\). If \(\defined\widetilde{\delta}(p, a)\), the reasoning is the same as in the base case. Suppose \(\neg\defined\widetilde{\delta}(p, a)\). Then \(\widetilde{\delta}_{\widetilde{f}}(p, a) = \widetilde{\delta}_{\widetilde{f}}(\widetilde{f}(p), a)\) and by the inductive hypothesis there exists a unique path \(\pi\) in \(\mathcal{F}\) from \(\widetilde{f}(p)\) to \(q\) with \(l(\pi) = a\) and \(w(\pi) = \widetilde{\lambda}_{\widetilde{f}}(\widetilde{f}(p), a)\). By \autoref{cloned-states-graph}, there exists an edge \(e \coloneqq \langle p, \langle \varepsilon, \widetilde{\varphi}(p) \rangle, \langle p, \widetilde{f}(p) \rangle \rangle\) in \(E\).
		
		If the length of \(\pi\) is \(1\), i.e. \(\pi = \langle \widetilde{f}(p), \langle a, \widetilde{\lambda}_{\widetilde{f}}(\widetilde{f}(p), a) \rangle, q \rangle\), then by \autoref{cloned-states-graph} there exists \(\langle \widetilde{f}(p), a, q \rangle \in \widetilde{\delta}\). Since \(\neg\defined\widetilde{\delta}(p, a)\), \(e' \coloneqq \langle \langle p, \widetilde{f}(p) \rangle, \langle a, \widetilde{\lambda}(\widetilde{f}(p) \rangle, a), q \rangle\) is an edge in \(\mathcal{G}\). Therefore, \(e e'\) is the unique path in \(\mathcal{G}\) from \(p\) to \(q\) with \(l(e e') = \varepsilon a = a\) and \(w(e e') = \widetilde{\varphi}(p)\widetilde{\lambda}(\widetilde{f}(p), a) = \widetilde{\lambda}_{\widetilde{f}}(p, a)\).
		
		If the length of \(\pi\) is greater than \(1\), i.e. \(\pi = \langle \widetilde{f}(p), \langle \varepsilon, \widetilde{\varphi}(\widetilde{f}(p)) \rangle, \langle \widetilde{f}(p), r \rangle \rangle \pi'\), then \(e' \coloneqq \langle \langle p, \widetilde{f}(p) \rangle, \langle \varepsilon, \widetilde{\varphi}(\widetilde{f}(p))\rangle, \langle \widetilde{f}(p), r \rangle \rangle \in E\). Thus, \(e e' \pi'\) is the unique path in \(\mathcal{G}\) from \(p\) to \(q\) with \(l(e e' \pi') = \varepsilon \varepsilon a = a\) and \(w(e e' \pi ') = \widetilde{\varphi}(p)\widetilde{\varphi}(\widetilde{f}(p))\frac{\widetilde{\lambda}_{\widetilde{f}}(\widetilde{f}(p), a)}{\widetilde{\varphi}(\widetilde{f}(p))} = \widetilde{\lambda}_{\widetilde{f}}(p, a)\).
	\end{proof}
\end{proposition}

\begin{proposition}\label{path-to-delta-f}
	Let \(\pi\coloneqq \langle p_0, \langle l_1, \omega_1\rangle, p_1,  \langle l_2, \omega_2\rangle, \ldots, p_n \rangle\), where \(n > 0\), be a path in \(\mathcal{G}\) such that \((\forall 1 \le i < n)(p_i \notin Q)\), \(p_0, p_n \in Q\) and \(l(\pi) = a \in \Sigma\). Then
	\[\widetilde{\delta}_{\widetilde{f}}(p_0, a) = p_n \land \widetilde{\lambda}_{\widetilde{f}}(p_0, a) = w(\pi) \land n > 1 \implies \neg\defined\widetilde{\delta}(p_0, a).\]
	\begin{proof}
		We proceed by induction on \(n\).
		
		First, suppose \(n = 1\). Then by \autoref{cloned-states-graph}, there exists \(\langle p_0, a, p_1 \rangle \in \widetilde{\delta}\) and the path is \(\langle p_0, \langle a, \widetilde{\lambda}(p_0, a) \rangle, p_1 \rangle\). Thus, \(\widetilde{\delta}_{\widetilde{f}}(p_0, a) = \widetilde{\delta}(p_0, a) = p_1\) and \(\widetilde{\lambda}_{\widetilde{f}}(p_0, a) = \widetilde{\lambda}(p_0, a) = w(\pi)\).
		
		Now, suppose \(n > 1\), i.e. \(\pi = \langle p_0, \langle \varepsilon, \widetilde{\varphi}(p_0)\rangle, \langle p_0, \widetilde{f}(p_0) \rangle \rangle \pi'\). The first state in \(\pi'\) is \(\langle p_0, \widetilde{f}(p_0) \rangle\) but the edges of \(\widetilde{f}(p_0)\) are a superset of the edges of \(\langle p_0, \widetilde{f}(p_0) \rangle\), therefore we can consider the path \(\pi''\) which is the same as \(\pi'\) with the exception that the first node is substituted with \(\widetilde{f}(p_0)\).
		
		If \(n - 1 = 1\), then there exists \(\langle \widetilde{f}(p_0), a, p_n \rangle \in \widetilde{\delta}\), \(\neg\defined\widetilde{\delta}(p_0, a)\) and \(\pi'' = \langle \widetilde{f}(p_0), \langle a, \widetilde{\lambda}(\widetilde{f}(p_0), a)\rangle, p_n \rangle \). Therefore,
		\begin{align*}
			\widetilde{\delta}_{\widetilde{f}}(p_0, a) &= \widetilde{\delta}_{\widetilde{f}}(\widetilde{f}(p_0), a) = \widetilde{\delta}(\widetilde{f}(p_0), a) = p_n, \\
			\widetilde{\lambda}_{\widetilde{f}}(p_0, a) &= \widetilde{\varphi}(p_0) \widetilde{\lambda}_{\widetilde{f}}(\widetilde{f}(p_0), a) = \widetilde{\varphi}(p_0) \widetilde{\lambda}(\widetilde{f}(p_0), a) = w(\pi).
		\end{align*}
		
		If \(n - 1 > 1\), by the inductive hypothesis for \(\pi''\)
		\[\widetilde{\delta}_{\widetilde{f}}(\widetilde{f}(p_0), a) = p_n \land \widetilde{\lambda}_{\widetilde{f}}(\widetilde{f}(p_0), a) = \frac{w(\pi)}{\widetilde{\varphi}(p_0)} \land \neg\defined\widetilde{\delta}(\widetilde{f}(p_0), a).\]
		Since \(\mathcal{\widetilde{W}}\) is monotonic, \(\neg\defined\widetilde{\delta}(p_0, a)\) and
		\begin{align*}
			\widetilde{\delta}_{\widetilde{f}}(p_0, a) &= \widetilde{\delta}_{\widetilde{f}}(\widetilde{f}(p_0), a) = p_n, \\
			\widetilde{\lambda}_{\widetilde{f}}(p_0, a) &= \widetilde{\varphi}(p_0) \widetilde{\lambda}_{\widetilde{f}}(\widetilde{f}(p_0), a) = \widetilde{\varphi}(p_0) \frac{w(\pi)}{\widetilde{\varphi}(p_0)} = w(\pi).
		\end{align*}
	\end{proof}
\end{proposition}

\begin{proposition}\label{delta-f-star-and-path}
	Let \(p, q \in Q\). Then 
	\begin{enumerate}
		\item for every \(\alpha \in \Sigma^*\) if \(\widetilde{\delta}_{\widetilde{f}}^*(p, \alpha) = q\), then there exists a unique path \(\pi\) in \(\mathcal{G}\) from \(p\) to \(q\) with \(l(\pi) = \alpha\) and \(w(\pi) = \widetilde{\lambda}_{\widetilde{f}}^*(p, \alpha)\);
		\item If there exists a path \(\pi\) in \(\mathcal{G}\) from \(p\) to \(q\) then \(\widetilde{\delta}_{\widetilde{f}}^*(p, l(\pi)) = q\) and \(\widetilde{\lambda}_{\widetilde{f}}^*(p, l(\pi)) = w(\pi)\).
	\end{enumerate}
	\begin{proof}
		Follows from \autoref{delta-f-to-path} and \autoref{path-to-delta-f} with a straightforward induction.
	\end{proof}
\end{proposition}

\begin{definition}\label{tilde-g-def}
	Let \(\widetilde{\mathcal{G}} \coloneqq (V \cup \{x\}, E^{rev} \cup \{\langle x, \langle \varepsilon, \widetilde{\rho}(q)\rangle, q \rangle \mid q \in Q_{s_1}\})\), where \(x \notin V\) is a new vertex. 
\end{definition}

\begin{remark}
	Since \(\widetilde{\rho}(q) = 1\) for \(q \in Q_{s_1}\), the edges in \(\widetilde{\mathcal{G}}\) from \(x\) are actually \(\{x\} \times \{\langle \varepsilon, 1 \rangle\} \times Q_{s_1}\).
\end{remark}

The use of \(\widetilde{G}\) is not necessary for the weight pushing in \(\mathcal{R}^+\), however it is essential in the \(\mathcal{R}^{\max}\) semiring described in the following section.

\begin{proposition}\label{tilde-g}
	\begin{enumerate}
		\item[]
		\item For every \(q \in Q\), \(S_{\widetilde{\mathcal{W}}}(q)\) is the sum of the paths in \(\widetilde{\mathcal{G}}\) from \(x\) to \(q\).
		\item If \(\mathcal{V}\) is acyclic, \(\widetilde{\mathcal{G}}\) is also acyclic.
	\end{enumerate}
	\begin{proof}
		\begin{enumerate}
		\item[]
		\item For a graph \(G \coloneqq \langle V, E \rangle\) we use \(\Pi_G(v, U)\) to denote the set of all paths in \(G\) from \(v\) to a vertex in \(U\). Analogously, \(\Pi_G(U, v)\) is the set of all paths in \(G\) from a vertex in \(U\) to \(v\).
    		\begin{align*}
    			S_{\widetilde{\mathcal{W}}}(q) &= \sum_{\alpha \in \Dom(\mathcal{O}_{\widetilde{\mathcal{W}}}^q)} \mathcal{O}_{\widetilde{\mathcal{W}}}^q(\alpha) \\
    			&= \sum_{\alpha \in \Dom(\mathcal{O}_{\widetilde{\mathcal{W}}}^q)} \widetilde{\lambda}_{\widetilde{f}}^*(q, \alpha) \\
    			&= \sum_{\substack{\alpha \in \Sigma^* \colon \\ \widetilde{\delta}_{\widetilde{f}}^*(q, \alpha) \in Q_{s_1}}} \widetilde{\lambda}_{\widetilde{f}}^*(q, \alpha) \\
    			&\stackrel{\mathclap{\normalfont\mbox{\tiny{\autoref{delta-f-star-and-path}}}}}{=} \sum_{\pi \in \Pi_\mathcal{G}(q, Q_{s_1})} w(\pi) \\
    			&= \sum_{\pi \in \Pi_\mathcal{\widetilde{G}}(Q_{s_1}, q)} w(\pi) \\
    			&= \sum_{\pi \in \Pi_\mathcal{\widetilde{G}}(x, q)} w(\pi).
    		\end{align*}
		\item We will show that \(\mathcal{G}\) is acyclic, which implies that \(\widetilde{\mathcal{G}}\) is also acyclic. Let \(\langle v_0, \langle l_1, \omega_1 \rangle, v_1, \langle l_2, \omega_2\rangle, \ldots, \omega_n, v_n \rangle\), \(n > 0\) be a path in \(\mathcal{G}\), such that \(v_n = v_0\). For \(0 \le i < n\) we known that \(v_i \notin Q_{s_1}\) because the states in \(Q_{s_1}\) have no outgoing transitions in \(\widetilde{\mathcal{W}}\).
		
		Suppose \((\exists 0 \le i < n)(\langle v_i, v_{i+1} \rangle \in \widetilde{f})\). Then \(\level_{\widetilde{f}}(v_{i+1}) < \level_{\widetilde{f}}(v_i)\) and \(\Proj_1(v_i) = \Proj_1(v_{i+1}) \in Q_\omega^l\) for some \(\omega \in \Omega\) (otherwise, \(v_i\) wouldn't have a failure transition). As long as the first coordinate of the vertex stays in \(Q_\omega^l\) the level of the vertex does not increase because either the second coordinate does not change, or a failure transition is followed. The level of the vertices in the cycle should increase for some \(j > i + 1\). Thus, for the smallest such \(j\) we have that \(\Proj_1(v_{j-1}) \in Q_\omega^l\) and \(\Proj_1(v_j) \in Q_\omega^r\). From \(v_j\) the only reachable vertex with first coordinate in \(Q_\omega^l\) is from \(Q_{s_1}\). But the cycle does not contain such vertices, which contradicts our assumption that the cycle contains a failure transition.
		
		Suppose \(\neg(\exists 0 \le i < n)(\langle v_i, v_{i+1} \rangle \in \widetilde{f})\). Then the first coordinates of the vertices in the cycle constitute a cycle in \(\mathcal{V}\). This is a contradiction because \(\mathcal{V}\) is acyclic.
		
		Therefore, \(\mathcal{G}\) and thus \(\widetilde{\mathcal{G}}\) are acyclic.
		\end{enumerate}
	\end{proof}
\end{proposition}

\begin{remark}
	\autoref{tilde-g} implies that to find the sums \(S_\mathcal{W}(q)\) for every \(q \in Q\) it suffices to compute the sums of the paths in \(\widetilde{\mathcal{G}}\) from \(x\) to every \(q\). This can be achieved with dynamic programming in linear time with respect to the size of \(\widetilde{\mathcal{G}}\).
\end{remark}

\subsection{Weight Pushing in \(\mathcal{R}^{\max}\)}

Let \(\mathcal{S} = \mathcal{R}^{\max}\) and \(\mathcal{G}\) be the graph from \autoref{cloned-states-graph} constructed from the failure transducer \(\mathcal{W}\). \autoref{delta-f-to-path}, \autoref{path-to-delta-f} and \autoref{delta-f-star-and-path} also hold in this semiring. Let \(\widetilde{\mathcal{G}}\) be the graph from \autoref{tilde-g-def} constructed from \(\mathcal{G}\) and \(\mathcal{W}\). Similar to \autoref{tilde-g}, in this semiring \(S_\mathcal{W}(q)\) is equal to the maximum of the paths in \(\widetilde{\mathcal{G}}\) from \(x\) to \(q\). We consider the isomorphism \(-log\) between the \(\mathcal{R}^{\max}\) semiring and the \(\langle \mathbb{R} \cup \{+\infty\}, \min, +, +\infty, 0 \rangle\) semiring. If we consider the graph \(-\log(\widetilde{\mathcal{G}})\) (the isomorphic image of \(\widetilde{\mathcal{G}}\)) with the same vertices as \(\widetilde{\mathcal{G}}\) and in which \(\langle v_1, \langle l, \omega \rangle, v_2 \rangle\) is an edge in \(\widetilde{\mathcal{G}}\) if and only if \(\langle v_1, \langle l, -\log(\omega)\rangle, v_2 \rangle\) is an edge in \(-\log(\widetilde{\mathcal{G}})\), then \(S_\mathcal{W}(q)\) will be equal to \(\exp(-\omega_q)\), where \(\omega_q\) is the weight of the shortest path in \(-\log(\widetilde{\mathcal{G}})\) from \(x\) to \(q\). If all weights in \(-\log(\widetilde{\mathcal{G}})\) are non-negative, then Dijkstra's algorithm can be used to compute the shortest paths from \(x\) to every node \(q\).

If \(-log(\widetilde{\mathcal{G}})\) has negative weights on some edges it can be shown that they correspond to failure transitions in the \(\mathcal{W}\) failure transducer. Since \(\mathcal{W}\) has no failure cycles, weight-pushing techniques similar to those in the previous section can be applied in order to make the weights non-negative. Afterwards, Dijkstra's algorithm can again be used to compute the shortest paths from \(x\) to every node \(q\). The detailed presentation of this method, however, is outside the scope of this paper.

\section{Acknowledgements}

The research presented in this paper is partially funded by the Bulgarian Ministry of Education and Science via grant DO1-200/2018 ``Electronic healthcare in Bulgaria'' (e-Zdrave) and  grant DO1-205/2018 ``Information and Communication Technologies for Unified Digital Market in Science, Education and Security''.

\bibliographystyle{splncs04}
\bibliography{references}

\end{document}